\documentclass[preprint,superscriptaddress,nofootinbib]{revtex4-1}
\usepackage{hyperref}
\usepackage{bbm}
\usepackage{amsfonts}
\usepackage{mathrsfs}

\usepackage{amsmath,amssymb}

\usepackage{epsfig}
\usepackage{graphicx}               
\usepackage{url}
\usepackage{hyperref}
\usepackage{float}
\usepackage{pstricks}
\usepackage{color}

\newcommand{\nc}{\newcommand}

\nc{\beq}{\begin{equation}}
\nc{\eeq}{\end{equation}}
\nc{\beqa}{\begin{eqnarray}}
\nc{\eeqa}{\end{eqnarray}}
\nc{\bea}{\begin{eqnarray}}
\nc{\eea}{\end{eqnarray}}
\nc{\barray}{\begin{eqnarray}}
\nc{\earray}{\end{eqnarray}}
\nc{\barrayn}{\begin{eqnarray*}}
\nc{\earrayn}{\end{eqnarray*}}
\nc{\ra}{\rightarrow}
\newcommand{\lsim}{\!\mathrel{\hbox{\rlap{\lower.55ex \hbox{$\sim$}} \kern-.34em \raise.4ex \hbox{$<$}}}}
\newcommand{\gsim}{\!\mathrel{\hbox{\rlap{\lower.55ex \hbox{$\sim$}} \kern-.34em \raise.4ex \hbox{$>$}}}}
\nc{\Tr}{{\rm Tr}}
\nc{\slsh}{\slash\hspace*{-0.22cm}}

\def\be{\begin{equation}}
\def\ee{\end{equation}}
\def\bea{\begin{eqnarray}}
\def\eea{\end{eqnarray}}

\nc{\infinity}{\infty}
\nc{\mc}{\mathcal}
\nc{\M}{\mathcal{M}}

\def\to{\rightarrow}

\newcommand\Bp{\tilde{B}'}
\newcommand\bino{\tilde{B}}
\newcommand\mBp{m_{\Bp}}

\begin{document}

\title{\Huge{Gaugino Portal Baryogenesis}
}

\vskip2cm

\author{Aaron Pierce}
\affiliation{Leinweber Center for Theoretical Physics, Department of Physics\\University of Michigan, Ann Arbor, MI 48109, USA}
\author{Bibhushan Shakya}
\affiliation{Santa Cruz Institute for Particle Physics, 1156 High St., Santa Cruz, CA 95064, USA}
\affiliation{Department of Physics, 1156 High St., University of California Santa Cruz, Santa Cruz, CA 95064, USA}
\affiliation{Department of Physics, University of Cincinnati, Cincinnati, OH 45221, USA}
\affiliation{Leinweber Center for Theoretical Physics, Department of Physics\\University of Michigan, Ann Arbor, MI 48109, USA}

\begin{abstract}

We study baryogenesis via a gaugino portal, the supersymmetric counterpart to the widely studied kinetic mixing portal, to a hidden sector. CP and baryon number violating decays of a hidden sector gaugino into the visible sector can produce the observed baryon asymmetry of the Universe. The tiny portal coupling is crucial in
producing late out-of-equilibrium decays, after washout processes that can erase the asymmetry have
gone out of equilibrium. We study this mechanism within various scenarios, including freeze-in or freeze-out of the hidden gaugino, as well as extended frameworks where the hidden sector contains a weakly interacting massive particle (WIMP) dark matter candidate. 
This mechanism can produce the desired asymmetry over a wide range of mass scales, including for hidden gaugino masses as low as 10 GeV.  We also discuss possible related signals with direct collider searches, at low energy experiments, and in dark matter direct and indirect detection. 
\end{abstract}

\preprint{LCTP-19-01}

\maketitle

\tableofcontents

\section{Motivation}
\label{sec:motivation}

The preferred picture for the underlying theory of nature has changed in the past few years. R-parity preserving weak scale supersymmetry, which offered a natural solution to the hierarchy problem as well as a WIMP dark matter candidate, is now constrained by data from a wide variety of experiments, ranging from the Large Hadron Collider (LHC) to dark matter indirect and direct detection efforts. Despite its many appealing theoretical features, supersymmetry, if realized in nature -- as a vestige of a theory of quantum gravity, or as a partial solution to the hierarchy problem -- might therefore be neither R-parity preserving nor at the weak scale. 

There has also been growing interest in the exploration of hidden or dark sectors, which can easily arise, for example, in string theories. Such sectors may communicate with our visible sector via one of the renormalizable portal interactions, giving rise to a rich array of phenomenological possibilities (See \cite{Alexander:2016aln} and references therein.). The most extensively studied of these is the kinetic mixing portal \cite{Holdom:1985ag}, where a gauged $U(1)'$ in a hidden sector has kinetic mixing with the Standard Model (SM) hypercharge $U(1)_Y$ \cite{Langacker:2008yv}, resulting in a mixing between the corresponding gauge bosons. 

In this paper, we explore how a hidden sector coupled to the Minimal Supersymmetric Standard Model (MSSM) via a kinetic mixing portal \cite{Dienes:1996zr} may address one of the outstanding problems of particle physics and cosmology, the origin of the observed excess of matter over antimatter in our Universe. Supersymmetry can readily satisfy the Sakharov conditions \cite{Sakharov:1967dj} necessary for baryogenesis: baryon number violation (from R-parity violating (RPV) operators), C and CP violation (from soft terms after supersymmetry breaking), and out of equilibrium interactions (from late decays of supersymmetric particles). In a supersymmetric scenario, kinetic mixing can occur between $U(1)$ gauge groups at the level of the supersymmetric field strengths, resulting in a gaugino mixing portal in addition to the more familiar gauge boson kinetic mixing portal.  Moreover, there is a possibility for mass mixing between the gauginos.  In this paper, we make use of late decays of the hidden sector gaugino into the visible sector via these mixings to produce the observed baryon asymmetry of the Universe. 

Within the MSSM, the most challenging aspect of implementing baryogenesis via RPV decays of neutralinos \cite{Dimopoulos:1987rk,Claudson:1983js,Sorbello:2013xwa,Cui:2012jh,Cui:2013bta,Arcadi:2015ffa,Barbier:2004ez} is to ensure that the decays occur sufficiently late as to avoid washout of the produced asymmetry from baryon number violating inverse decay and scattering processes without suppressing the production mechanism.  In low scale baryogenesis mechanisms one must also be wary of inducing a too-large CP violation that would be visible in electric dipole moment (EDM) experiments, which are now strongly constrained by measurements (For recent discussions of the implications of EDM bounds for supersymmetric theories, see \cite{Altmannshofer:2013lfa,Cesarotti:2018huy}.). In this paper, we demonstrate that the hidden sector implementation offers several novel features: the small portal mixing between the two sectors naturally provides the means to address the above problems, enabling low scale baryogenesis with rich phenomenology. 

Furthermore, while R-parity violating supersymmetry has an unstable lightest supersymmetric particle (LSP) and therefore no dark matter candidate\,\footnote{If R-parity is unbroken, the presence of the gaugino portal nevertheless carries interesting cosmological implications for dark matter, see \cite{Ibarra:2008kn,Arvanitaki:2009hb,Acharya:2016fge}.} (except possibly a long-lived gravitino, see, e.g., \cite{Arcadi:2015ffa}), the hidden/dark sector can contain additional particles that account for the dark matter. In this paper, we will study both a minimal hidden sector (where the only hidden sector particle relevant for baryogenesis is the hidden sector gaugino) and an extended hidden sector containing a WIMP dark matter particle, which also impacts the process of baryogenesis.

\section{Framework}
\label{sec:framework}

Our framework consists of the MSSM extended by two key ingredients: baryon number violation via an RPV coupling, and a kinetic mixing portal to a $U(1)'$ symmetry residing in a hidden sector.  

For baryon number violation, we add to the MSSM superpotential the following RPV term
\beq
W_{RPV}=\lambda''_{ijk}U_i^c D_j^c D_k^c.
\label{eq:rpv}
\eeq
We set other RPV terms, which break lepton number and can induce proton decay, to zero. The above RPV coupling $\lambda''_{ijk}$ is constrained by various measurements\,\footnote{For a somewhat old but comprehensive discussion of RPV supersymmetry and constraints, see \cite{Barbier:2004ez}.}, depending on the flavor indices $(i,j,k)$. 
In this paper, for simplicity we work with a single $\lambda''_{ijk}$ without specifying the flavor indices.

Regarding the field content of the hidden sector, two variations are worth studying. One can consider a minimal setup where all of the relevant hidden sector gaugino phenomenology arises from the portal coupling to the visible sector. Alternatively, additional particles in the hidden sector may endow the hidden gaugino with additional interactions, which can affect baryogenesis. We now outline these two possibilities in turn, before returning to discuss their phenomenology.

\subsection{Minimal Hidden Sector}

Kinetic mixing between two abelian gauge field strength superfields has been studied in \cite{Dienes:1996zr,Ibarra:2008kn,Arvanitaki:2009hb}; the relevant Lagrangian is
\beq
\mathcal{L}=\frac{1}{32}\int d^2\theta \{W_v W_h + W_v W_h-2\eta W_v W_h\}
\eeq
where we use the notation $W_v$ and $W_h$ to denote the chiral gauge field strength superfields for the visible sector SM hypercharge $U(1)_Y$ and the hidden sector $U(1)'$ gauge symmetries respectively. Here, $\eta$ is the kinetic mixing portal coupling between the two fields. Typically, $\eta\sim10^{-3}$ if the mixing  is induced at one loop by chiral superfields charged under both gauge groups, but much smaller values are possible in, e.g., compactifications of heterotic and type II strings \cite{Dienes:1996zr,Blumenhagen:2006ux,Lust:2003ky,Abel:2003ue,Berg:2004ek,Abel:2006qt,Abel:2008ai}. The gauge kinetic terms can be made canonical by shifting the visible and hidden sector vector superfields $V_v$ and $V_h$ (See \cite{Dienes:1996zr,Ibarra:2008kn,Arvanitaki:2009hb} for details.). If the two gauginos have Majorana masses $M_v, M_h$, these shifts result in a small Dirac mass term $\sim\eta M_h$ between the two gauginos. Depending on the underlying model of supersymmetry breaking, an additional primordial Dirac mass term might be present, which may well be of the same order. While one could, in principle, proceed by eliminating the kinetic mixing via field redefinition, followed by diagonalizing the mass matrix,  in this paper we work with a ``simplified model" setup for the gaugino sector that captures the same effects:
\beq
\mathcal{L}\supset -\frac{1}{2}(\mBp \Bp\Bp+m_{\bino}\bino\bino+\epsilon\,\mBp \Bp\bino),
\label{binomasses}
\eeq  
where we $\textit{define}$ $\epsilon$ to be the mixing between the hidden gaugino, $\Bp$, and the MSSM bino, $\bino$, but do not imagine there is otherwise any kinetic mixing in the gaugino sector\,\footnote{There are also  ${\mathcal O}(\epsilon)$ couplings between the hidden gauge boson and the visible sector, not relevant for gaugino decays but important, for example, for allowing the decay of the hidden gauge bosons.}. We expect $\epsilon\sim\eta$ unless there are strong cancellations between the various contributions to the Dirac mass term between the $\Bp$ and the $\bino$ (See \cite{Ibarra:2008kn} for related discussion.). Upon diagonalization of the neutralino mass matrix, this mass mixing will induce ${\mathcal O}(\epsilon)$ couplings between the (primarily) hidden neutralino and the visible sector. We assume $\mBp\sim m_{\bino}$, as would be the case, e.g., with gravity mediation of supersymmetry breaking to both sectors.

The MSSM bino $\bino$ can mix with the other MSSM neutralinos, the wino and the Higgsinos. Such mixing is suppressed by ${\mathcal O}(M_{Z}/\mu)$, with $\mu$ the Higgsino mass term. If the hidden sector $U(1)'$ is spontaneously broken by a hidden Higgs vacuum expectation value (vev), one analogously also has mixing of the hidden sector gaugino with the hidden sector Higgsinos (for simplicity, we assume two Higgsinos for the sake of anomaly cancellation, as in the MSSM), which can also be suppressed by raising the hidden Higgsino mass term $\mu^{\prime}$. In this paper, for simplicity, we assume $\mu' \gg M_{Z'}$ and $\mu\gg M_{Z}$. This decouples the remaining neutralinos, and the only relevant gauginos for our purposes are the MSSM bino $\tilde{B}$ and the hidden sector gaugino $\tilde{B'}$. 

From now on, we will use the notation $\Bp, \bino$ to denote the mass eigenstates, which contain $\epsilon$ admixtures of the gauge eigenstate of the opposite sector. Crucially, we also assume that the two mass terms carry a relative phase: Im$(\mBp m_{\bino}^*) \neq 0$. This will be the source of CP violation necessary for baryogenesis. Finally, we assume that $\bino$ is the LSP, so it can only decay via the RPV coupling.   

In addition to the hidden gaugino $\Bp$, the hidden sector contains, at minimum, Higgs bosons $h'$, Higgsinos $\tilde{H}'$, and the dark gauge boson $Z'$. These particles decay much faster than the $\Bp$:  For the mass hierarchy $m_{\tilde{H}'} > m_{h'} > \mBp$, the dark Higgsino decays as $\tilde{H}'\to h' \Bp$, the dark Higgs boson decays as $h'\to \Bp\Bp$ via small but non-vanishing $\Bp-\tilde{H}'$ mixing, while the $Z'$ boson undergoes 2-body decays into the SM via a kinetic mixing term. The $\Bp$ interactions with both the $Z'$ and the $h'$ vanish as the Higgsinos $\tilde{H}'$ are decoupled. Therefore, the presence of these additional particles in the hidden sector is cosmologically irrelevant, and all relevant interactions of the $\Bp$ arise via its $\epsilon$ mixing with the visible sector bino $\bino$. 

\subsection{Extended Hidden Sector with WIMP Dark Matter}
\label{sec:extended}

The presence of additional particles in the hidden sector is an attractive possibility from the point of view of dark matter, since the traditional dark matter candidate, the visible sector LSP, is unstable in the absence of R-parity\,\footnote{With RPV, a gravitino LSP with a sufficiently long lifetime can account for dark matter, see e.g. \cite{Arcadi:2015ffa}.}. While hidden sectors allow for a wide variety of possibilities in terms of additional field content, there are two important considerations to keep in mind:
(1) additional field content should not introduce new $\Bp$ decay channels into the hidden sector, as these would dominate over the $\epsilon$ suppressed decay process generating the baryon asymmetry, and (2) additional field content should preserve the cancellation of anomalies related to the gauged $U(1)'$. There are two straightforward ways to ensure anomaly cancellation: adding a $U(1)'$ singlet, or adding vector-like fields.

Adding a singlet opens up possible additional portal interactions between the hidden and visible fields. While such setups admit interesting possibilities, these are not directly related to the baryogenesis process of interest. Therefore, we consider instead a vector-like fermion $X$, with mass $m_X$, charged under the $U(1)'$ with gauge coupling $g_D$, which is stable under a $\mathbb{Z}_2$ symmetry and therefore a good dark matter candidate.  In particular, for weak scale $m_{X}>M_{Z'}$, it can realize the correct abundance via the ``WIMP miracle" through the freeze-out process $X\bar{X} \to Z'Z'$, where $Z'$ is the hidden sector gauge boson. Furthermore, while it does not introduce any new $\Bp$ decay channels, it does introduce additional Yukawa couplings for the $\Bp$ proportional to the hidden gauge coupling, involving its scalar superpartner $\tilde{X}$. As we will see in the following sections, this can affect the cosmological history of the hidden gaugino and therefore the production of the baryon asymmetry, offering an interesting variation to the minimal setup discussed above.

\section{Calculation of Baryon Asymmetry}
\label{sec:baryogenesis}

\begin{figure}[t]
\includegraphics[width=4.5in]{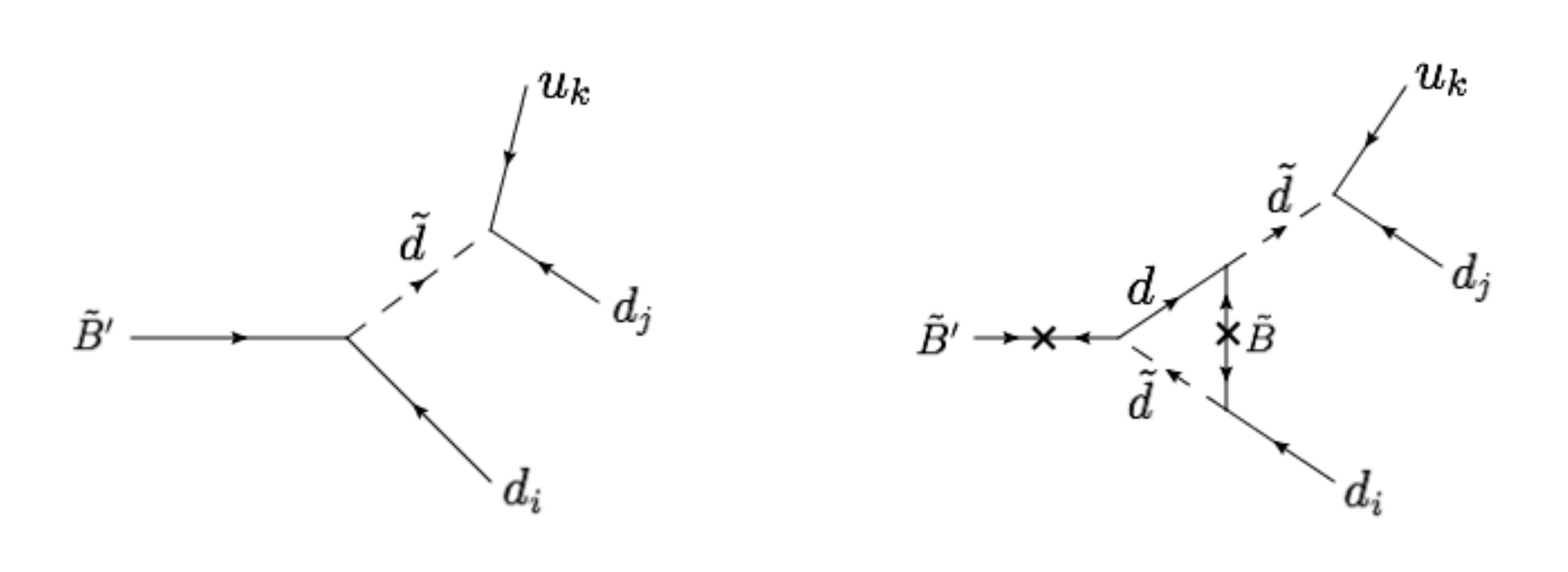}
\caption{\label{fig:decays} Tree and loop level decays of the hidden gaugino $\Bp$ that interfere to produce a baryon asymmetry. The crosses in the second diagram represent mass insertions, necessary to introduce the phases on the two gaugino masses that give rise to the CP asymmetry in the interference term.}
\end{figure}

The framework described in the previous section contains all the necessary ingredients for baryogenesis via the early Universe production and decay of the hidden gaugino $\Bp$ into SM fermions in the early Universe. The necessary ingredients are the Sakharov conditions \cite{Sakharov:1967dj} of baryon number violation, C and CP violation, and out of equilibrium interactions. Baryon number violation is made possible by the RPV interaction in Eq.~(\ref{eq:rpv}). The out of equilibrium condition is provided by late $\Bp$ decays. Finally, CP violation arises from the interference of the tree and loop level decays shown in Fig.\,\ref{fig:decays} via the non-trivial phase in the gaugino masses.

The baryon asymmetry generated with this setup can be calculated as:
\be
Y_{BA} \equiv \frac{n_{b} - n_{\bar{b}}}{s} = \epsilon_{CP}\, Y_{\Bp}\,W_I,
\ee
where $Y_{\Bp}= n_{\Bp}/s$ is the freeze-out abundance of the $\Bp$, and
\be
\epsilon_{CP} \equiv \frac{\Delta \Gamma_{BV}}{\Gamma_{\Bp}} \equiv \frac{\Gamma_{\Bp\to u_id_jd_k}-\Gamma_{\Bp\to\bar{u_i}\bar{d_j}\bar{d_k}}}{\Gamma_{\Bp}},
\ee
(where $\Gamma_{\Bp}$ is the total decay width of the $\Bp$) represents the fraction of $\Bp$ decays that produce a baryon asymmetry. $W_I$ incorporates the washout or dilution of the asymmetry from subsequent processes. Note that, due to the long lifetime of the $\Bp$, its production/freeze-out and decays occur at different epochs, and the related quantities, $Y_{\Bp}$ and $\epsilon_{CP}$ respectively, can be calculated independently. In the remainder of this section, we discuss the calculation of each of the above contributions. 

\subsection{Calculation of $\epsilon_{CP}$}

Since the decay processes shown in Fig.\,\ref{fig:decays} depend on $\epsilon$ as well as $\lambda''$, the decay can have a long lifetime, easily satisfying the out of equilibrium condition. 

Here, it is important to note that there exists another decay channel in addition to those shown in Fig.\,\ref{fig:decays}. As a consequence of the Nanopoulos-Weinberg theorem \cite{Nanopoulos:1979gx}, in order for the generated CP asymmetry to be first order in the RPV coupling, the $\Bp$ must have an additional decay channel when the RPV coupling is set to zero (See~\cite{Sorbello:2013xwa} for a detailed discussion.). This is realized for $m_{\bino}<\mBp$, \footnote{One can instead also have a lighter gluino or wino, as considered in \cite{Arcadi:2015ffa,Cui:2013bta}; we do not consider these possibilities in this paper and focus instead on the lighter bino, which is both novel from the hidden sector implementation as well as the least constrained experimentally.} for which the $\Bp$ has the additional decay channel $\Bp\to \bino f\bar{f}$, which is independent of the RPV coupling $\lambda''$, but notably depends on $\epsilon$. Including this channel, the decay widths relevant for the calculation of the baryon asymmetry are \cite{Arcadi:2015ffa,Cui:2013bta,Baldes:2014rda}
\bea
\Delta\Gamma_{\Bp}=\Gamma (\Bp\to udd)-\Gamma(\Bp\to\bar{u}\bar{d}\bar{d})&=&\frac{3\,\epsilon^2\lambda''^2\alpha_1^2}{256\pi^2}Im[e^{2 i\phi}]\frac{\mBp^6\,m_{\bino}}{m_0^6}\, f_2(m_{\bino}^2/\mBp^2),\\
\Gamma (\Bp\to udd+\bar{u}\bar{d}\bar{d})&=&\frac{3\,\epsilon^2\lambda''^2\alpha_1}{128\pi^2}\frac{\mBp^5}{m_0^4},\\
\Gamma (\Bp\to \bino f\bar{f})&=&\frac{\epsilon^2\alpha_1^2}{64\pi}\frac{\mBp^5}{m_0^4}\,f_2(m_{\bino}^2/\mBp^2).
\eea
where $f_2(x)=1-8x+8x^3-x^4-12x^2$log$(x)$, and $m_0$ is the sfermion mass scale. $\phi=\phi_{\bino}-\phi_{\Bp}$ is the relative phase between the two gaugino masses, and from here on we set it to its maximum value, Im$[e^{2i\theta}]=1$, which maximizes the baryon asymmetry.  There are in principle additional decay channels, such as $\Bp\to\bino (h/Z)$, from the suppressed bino-Higgsino mixing.  For sufficiently large values of $|\mu| > \frac{8\pi}{g_1} \left(\frac{m_0}{\mBp}\right)^2\!m_Z$, which we assume to be the case, this contribution is negligible compared to the channels above.

With the above decay rates, we have
\be
\epsilon_{CP}=\frac{\Delta\Gamma_{\Bp}}{\Gamma_{\text{total}}}=\frac{3}{4\pi}\frac{ \lambda''^2 }{1+3 \lambda''^2/[2\pi\alpha_1 f_2(m_{\bino}^2/\mBp^2)]} \left(\frac{\mBp m_{\bino}}{m_0^2}\right)\,.
\label{eq:ecp}
\ee
The various suppression factors in this expression are intuitive. The $4\pi$ in the first denominator represents the loop factor suppression in the tree-loop interference relative to the tree level decay. The second fraction represents the additional coupling in the interference term. Finally, the $1/m_0^2$ factor in the third fraction can be understood from the presence of an additional sfermion propagator in the tree-loop interference diagram relative to the tree level process. Given the mass hierarchy $m_0>\mBp>m_{\bino}$, the above expression evaluates to $\lsim10^{-4}$. 

We now highlight some important aspects of the $\Bp$ decay. Suppose $\epsilon$ were not small.  Since the $\Bp\to \bino f\bar{f}$ channel is independent of the RPV coupling $\lambda''$, one cannot make the $\Bp$ long-lived simply by suppressing $\lambda''$. This could be accomplished, instead, by raising the sfermion mass scale $m_0$ relative to the gaugino masses (as occurs in split-supersymmetry scenarios \cite{Wells:2003tf,ArkaniHamed:2004fb,Giudice:2004tc,Wells:2004di}), as implemented in the MSSM baryogenesis scenarios in \cite{Arcadi:2015ffa,Cui:2013bta}.  However, this has the effect of also suppressing $\epsilon_{CP}$ (see the final factor in Eq.~(\ref{eq:ecp})), therefore suppressing the production of the baryon asymmetry. The hidden sector implementation is qualitatively different and advantageous in this regard: the tiny portal coupling $\epsilon$ suppresses all decay rates, thereby making the $\Bp$ long-lived, but without an accompanying suppression of  $\epsilon_{CP}$.   

\subsection{Calculation of $Y_{\Bp}$: Minimal Hidden Sector}

In the minimal implementation, there are three distinct types of cosmological histories\,\footnote{For a detailed discussion of these various phases, see \cite{Hall:2009bx}.} determining the hidden gaugino abundance $Y_{\Bp}$ in the early Universe.

\begin{figure}[t]
\includegraphics[width=4.5in]{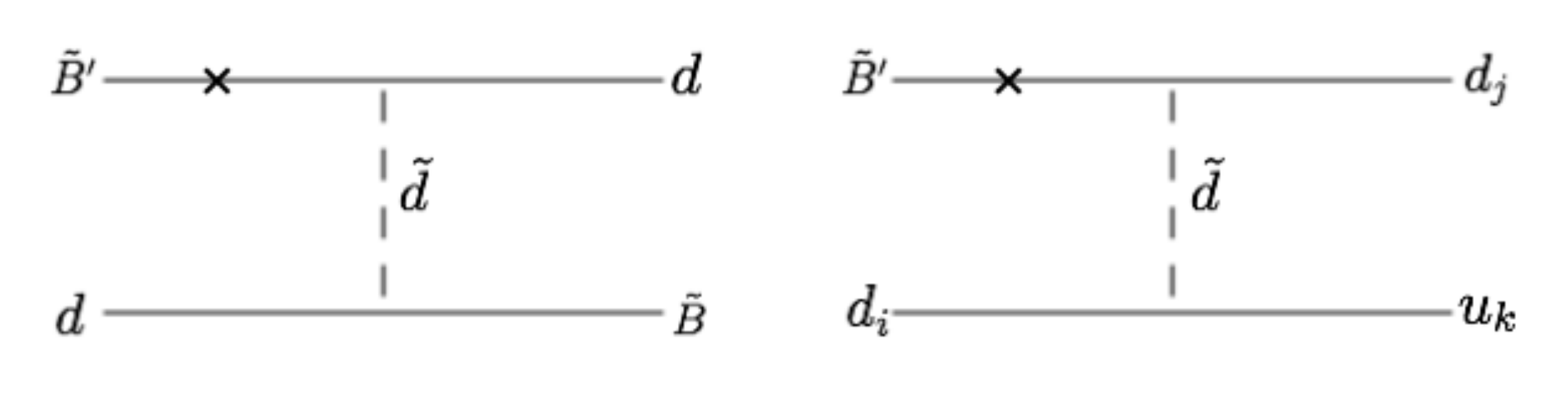}
\caption{\label{fig:annihilationm} Leading $\Bp$ interactions that determine its freeze-out in the minimal scenario. The crosses represent the $\epsilon$ mixing angle between the $\Bp$ and the $\bino$.}
\end{figure}

\textit{Conventional Freeze-Out:} If $\epsilon$ is sufficiently large that the $2\to 2$ scattering processes shown in Fig.\,\ref{fig:annihilationm} are rapid (compared to the Hubble rate) in the early Universe, the $\Bp$ thermalizes with the SM bath and subsequently undergoes conventional freezeout. While in equilibrium, its number density is given by \cite{Baldes:2014rda}
\beq
n_{eq}(\Bp)=\frac{ M^2 T}{\pi^2} K_2 (x),
\label{eq:eqabundance}
\eeq
where $x=\mBp/T$ and $K_2$ is the modified Bessel function of the second kind. The thermally averaged cross sections for the processes in Fig.\,\ref{fig:annihilationm} are \cite{Arcadi:2015ffa}
\bea
\langle\sigma v\rangle(\Bp u \to \bino \bar u)+\langle\sigma v\rangle(\Bp d \to \bino \bar d)&=&\frac{4\pi\epsilon^2\alpha_1^2}{36}\frac{\mBp^2}{m_0^4}\left(8\frac{K_4(x)}{K_2(x)}+1\right),\\
\langle\sigma v\rangle(\Bp u_k \to \bar d_i \bar d_j)+\langle\sigma v\rangle(\Bp d_i \to \bar u_k d_j d)&=&\frac{\epsilon^2\alpha_1\lambda''^2}{3}\frac{\mBp^2}{m_0^4}\left(5\frac{K_4(x)}{K_2(x)}+1\right).
\eea
In our studies, we calculate the freezeout abundance $Y_{\Bp}$ by numerically solving the Boltzmann equation for the $\Bp$ (for details, see \cite{Arcadi:2015ffa}) with the above interaction terms. 

\textit{Thermal abundance from decays:} For smaller values of $\epsilon$,  the $2\to 2$ scattering processes above are not sufficiently rapid to thermalize the $\Bp$. Nevertheless, an equilibrium abundance of the $\Bp$ (Eq.\,(\ref{eq:eqabundance})) can be achieved from decays such as $\tilde{f}\to\Bp f$ at $T>\mBp$. This scenario is realized for $\alpha_1\epsilon^2>m_0/M_{Pl}$, and results in a large yield $Y_{\Bp}\approx 5\times 10^{-3}$.

\textit{Freeze-in:} For even smaller values of $\epsilon$, $Y_{\Bp}$ never reaches thermal abundance, and a smaller abundance accumulates instead via the freeze-in process \cite{Hall:2009bx}. Of the feeble processes, the contribution from decays of sfermions while they are in equilibrium, $\tilde{f}\to\Bp f$, generally dominates over contributions from scattering processes in the IR, and contributes \cite{Hall:2009bx}
\beq
Y_{\Bp}\approx \frac{0.03\, \alpha_1\,\epsilon^2}{\pi^3}\frac{M_{Pl}}{m_0}.
\eeq

\subsection{Calculation of $Y_{\Bp}$: Extended Hidden Sector}

The presence of the vector-like fermion $X$ in the hidden sector could drastically change the picture. In this case, the $\Bp$ interacts with $X$ via the t-channel exchange of the sfermion $\tilde{X}$. This is analogous to the familiar case of the freeze-out of the MSSM bino with leptons via the exchange of a slepton mediator, and has the annihilation cross section \cite{ArkaniHamed:2006mb}: 
\beq
\sigma v (\Bp\Bp\to X \bar{X})\approx \frac{g_D^4 r(1+r)^2}{2\pi m_{\tilde{f}}'^2 x(1+r)^4},
\eeq
where $r=\mBp^2/m_{\tilde{f}'}^2$.  This cross section is independent of $\epsilon$ and can therefore dominate the early Universe dynamics of the $\Bp$ despite the Boltzmann suppressed abundance of the $\Bp$ at $T<\mBp$. If this interaction controls the freeze-out of the $\Bp$, the abundance $Y_{\Bp}$ can be estimated, in the limit  $r=\mBp^2/m_{\tilde{f}'}^2\ll 1$, as
\beq
Y_{\Bp}\approx 10^{-4}\left(\frac{0.1}{g_D}\right)^4\left(\frac{m_{\tilde{f}'}}{10\,\text{TeV}}\right)^4\left(\frac{\text{TeV}}{\mBp}\right)^3.
\label{eq:yextended}
\eeq

In our studies, we will focus on regions of parameter space where the WIMP-like freeze-out $X\bar{X}\to Z'Z'$ sets the correct relic density of dark matter, and $\Bp\Bp\to X \bar{X}$ determines the freeze-out abundance of the $\Bp$.  This requires not only $m_{Z'} < m_X$, but also $\epsilon$ large enough that the visible and hidden sectors are in equilibrium at some point in the early Universe, and  $\mBp>m_X$  so that the $\Bp$ freeze-out can occur at $T<\mBp$.  Note that requiring the correct dark matter abundance fixes a relation between $g_D$ and $m_X$: in the limit $m_{Z'}/m_X\ll 1$, we have $g_D\approx 0.5\sqrt{m_X/\text{TeV}}$.

\subsection{Washout and Dilution Effects}

In our framework, the main effects that can wash out the produced baryon asymmetry are the baryon number violating inverse decays or $2\to2$ annihilations involving the $\bino$, for which the rates are \cite{Arcadi:2015ffa}
\bea
\Gamma_{ID}&=&\frac{3\,\lambda''^2\alpha_1}{128\pi^2}\frac{\mBp^5}{m_0^4}x^2 K_2 (zx),\nonumber\\
\Gamma_{S}&=&\frac{\alpha_1\lambda''^2}{4\pi}\frac{\mBp^5}{m_0^4}\left(5\frac{K_4(x)}{K_2(x)}+1\right)K_2 (zx),
\label{eq:inverse}
\eea
where $z=m_{\bino}/\mBp$. Washout effects are important if the above processes have not gone out of equilibrium (i.e., are rapid compared to the Hubble rate) at the time of the $\Bp$ decay. However, even in this case, since the decay of the entire $\Bp$ ensemble is not instantaneous, the reduced $\Bp$ abundance $Y_{\Bp}\,e^{-\Gamma_{\Bp} t}$ remaining at the time the washout processes go out of equilibrium ($\Gamma_{ID}+\Gamma_{S} < H$) can still be sufficient to produce the desired baryon asymmetry.

Another effect that can suppress the baryon asymmetry is entropy dilution from the late decays of the $\Bp$. The $\Bp$ energy density redshifts slower than radiation for $T < \mBp$. Since the $\Bp$ is long-lived, its energy density can therefore grow to dominate the total energy density of the Universe at late times, when $Y_{\Bp} \mBp/T > 1$. The eventual decay of the $\Bp$ population injects this entropy into the thermal bath, diluting the abundance of the baryon asymmetry. This dilution factor can be estimated as \cite{Kolb:1990vq}
\beq
\label{eq:dilution}
\frac{S_{\text{after}}}{S_{\text{before}}}\approx1.83\,g_*^{1/4} Y_{\Bp}\frac{\mBp}{\sqrt{\Gamma_{\Bp} M_{Pl}}}\,.
\eeq
This significant entropy injection also raises the temperature of the thermal bath to 
\beq
\label{eq:reheatT}
T_{\text{after}}\approx 0.55 g_*^{-1/4}\sqrt{\Gamma_{\Bp} M_{Pl}},
\eeq
which affects the calculation of washout effects. We take these dilution and reheating effects into account where relevant.  

\section{Results and Discussion}
\label{sec:results}

In this section, we present the results of our calculation of the baryon asymmetry using the formalism described in the previous sections. For both the minimal and extended hidden sector setups, we perform numerical scans over the relevant parameter space, exploring regions that can produce the observed amount of baryon asymmetry in the Universe, $Y_{BA}\approx 10^{-11}$.  

\subsection{Minimal Hidden Sector}

In the minimal scenario, the free parameters are $\mBp,\,m_{\bino},\,m_0,\epsilon,$ and $\lambda''$.

\begin{figure}[t]
\includegraphics[width=3.8in]{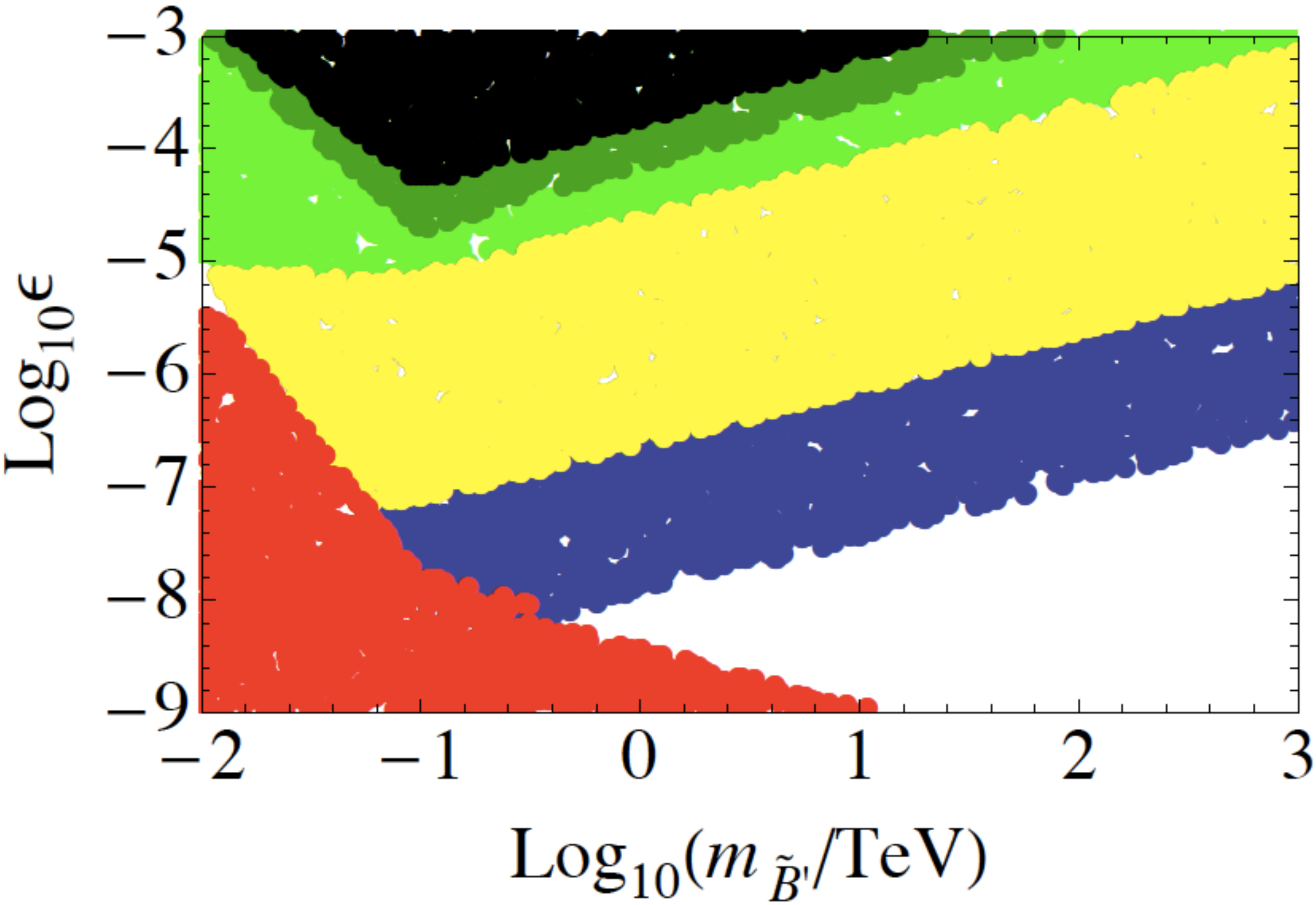}
\caption{\label{fig:fixedscan} Regions of parameter space where the desired amount of baryon asymmetry can be produced from cosmological histories where the $\Bp$ freezes out (green, yellow) or freezes in (blue), for $m_0=$\,max$(1$ TeV$,10\,\mBp),~m_{\bino}=0.3\,\mBp,$ and $\lambda''=0.1$. In the green region, the $\Bp$ thermalizes and freezes out via scattering processes (Fig.\,\ref{fig:annihilationm}), whereas in the yellow region the $\Bp$ reaches equilibrium abundance by virtue of the decay process $\tilde{f}\to\Bp f$. Dark green and black points denote regions where the washout processes (baryon number violating inverse decay and/or annihilation) are faster than Hubble at the time of $\Bp$ decay; in the dark green regions, a sufficient amount of baryon asymmetry survives despite partial washout, while in the black region the observed baryon asymmetry in the Universe cannot be matched. Red region denotes parameter space where the $\Bp$ decay occurs after BBN ($T=1$ MeV). }
\end{figure}

To understand the various factors at play, we first plot various regions of interest in the $\mBp-\epsilon$ plane in Fig.\,\ref{fig:fixedscan}, fixing the remaining parameters as specified in the caption. As discussed earlier, there are several classes of viable cosmological histories that produce the observed baryon asymmetry: the $\Bp$ can thermalize and freeze out via scattering processes (green), reach thermal abundance due to the decay process $\tilde{f}\to\Bp f$ (yellow), or build up a small abundance from freeze-in (blue). Other colors indicate problematic regions, where the $\Bp$ decay occurs after big bang nucleosynthesis (BBN), taken to be $T=1$ MeV (red), or where the baryon number violating inverse decay and/or annihilation washout processes suppress $Y_{BA}$ below the desired value (black). In the dark green region, washout processes are active at the temperature at which the $\Bp$ decays, but a sufficient amount of baryon asymmetry survives, as discussed earlier (See discussion after Eq.\,(\ref{eq:inverse}).). 
 
It is straightforward to understand the transitions between these regions as a function of $\epsilon$.
Extremely small $\epsilon$ leads to too-long lifetime (red). Increasing $\epsilon$ not only reduces the lifetime (white) but also leads to greater production of the $\Bp$ through the freeze-in process $\tilde{f}\to\Bp f$, eventually enabling sufficient production of the baryon asymmetry (blue). As $\epsilon$ is further increased, this decay process leads to a thermal abundance of the $\Bp$ from decay processes (yellow).  Eventually scattering processes (Fig.\,\ref{fig:annihilationm}) also become rapid, and control the freeze-out of the $\Bp$ (green). For even larger values of $\epsilon$, the decay lifetime can become so short that the decay occurs while washout processes are still active, partially erasing the baryon asymmetry (dark green) and eventually making it impossible to realize the desired abundance (black). 

The change in the behavior of the black and red regions below $\mBp=100$ GeV is due to our choice of $m_0=$ max$(1$ TeV$,10\,\mBp)$, in view of LHC bounds on sub-TeV squarks. We also note that, in the majority of the yellow and blue points, the $\Bp$ dominates the energy density of the Universe at the time of its decay, resulting in dilution of the baryon asymmetry (see discussion surrounding Eq.~(\ref{eq:dilution}),(\ref{eq:reheatT})); the final abundance can nevertheless match the observed value.

Next, we perform an extensive scan over parameter space covering the following ranges:
\begin{itemize}
\item 10 GeV $< \mBp < $ 10$^8$ GeV. 
\item $2  < \mBp/m_{\bino} < 1000$.
\item Log$_{10}(m_0/\mBp)$ between 0.5 and 5. Since there are LHC limits on sub-TeV squarks even for RPV supersymmetry \cite{Aaboud:2017dmy,Bardhan:2016gui,Li:2018qxr}, we additionally also enforce $m_0>1$ TeV.
\item $10^{-10}  < \epsilon <10^{-3}$. 
\item $10^{-4} < \lambda'' < 0.5$. 
\end{itemize}

\begin{figure}[t]
\includegraphics[width=4in]{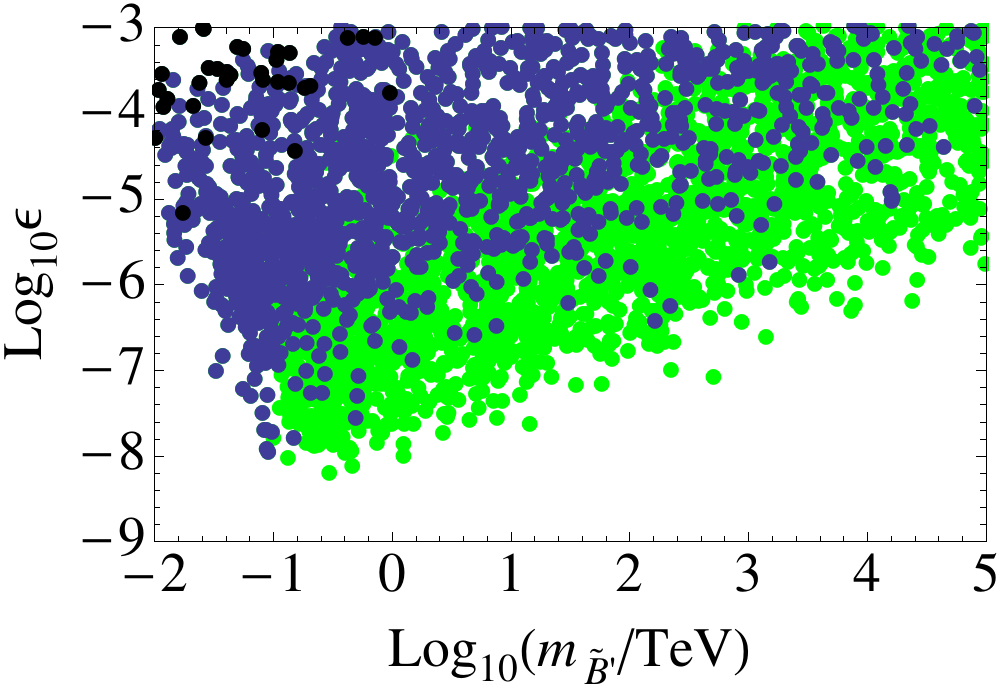}
\caption{\label{fig:fullscan} Parameter space that produces a sufficient amount of baryon asymmetry consistent with all constraints. Points are color coded according to $m_0/\mBp < 10$ (green), $10 < m_0/\mBp < 100$\,(blue), or $100 < m_0/\mBp < 1000$\,(black).}
\end{figure}

The distribution of points compatible with all constraints (sufficient amount of baryon asymmetry, the $\Bp$ decay before BBN, and washout processes inactive at the time of $\Bp$ decay) in the $\mBp-\epsilon$ plane is shown in Fig.\,\ref{fig:fullscan}. We see that the desired amount of baryon asymmetry can be produced over a wide range of parameter space, spanning several orders of magnitude in $\mBp$ and $\epsilon$. The three major constraints that bound this region are:  the requirement that the $\Bp$ decays before BBN (towards the bottom left of the figure), the requirement of sufficient $\Bp$ production (bottom right), and the absence of strong washout effects (top left). The smallest possible value of $\epsilon$ in the compatible region of parameter space is $\approx 10^{-8}$, for $\mBp\approx 100$ GeV. The desired baryon asymmetry can be generated for $\mBp$ as small as $10$ GeV (the observed patterns suggest that viable regions can be found even for GeV scale $\Bp$; however, for such low masses the details of the exact decay channel and phase space suppression might become relevant), extending to extremely heavy masses $(\,> 10^8$ GeV).   

The color coding represents various values of $m_0/\mBp$ (see caption), and suggests a correlation between large $m_0/\mBp$ and large $\epsilon$. This correlation is enforced by the need to keep various interaction rates sufficiently rapid to maintain sufficient production of the asymmetry and/or avoid constraints (decay before BBN, avoid washout of asymmetry). Various other parameter combinations are similarly constrained from their effect on $\epsilon_{CP}$ (see Eq.\,(\ref{eq:ecp})). For instance, from our scans we find that $\lambda''<10^{-3}$ cannot produce the desired amount of baryon asymmetry;  this can be understood from noting that $\lambda''<10^{-3}$ forces $\epsilon_{CP}\lsim 10^{-9}$ (see Eq.~(\ref{eq:ecp})), making it impossible to obtain a sufficiently large asymmetry since $Y_{\Bp}\lsim0.01$. Similar considerations likewise restrict $m_0/\mBp\lsim 1000$, as verified by the scan.

\subsection{Extended Hidden Sector}

We now study how the parameter space changes with the extended hidden sector, in particular, whether the baryogensis mechanism discussed above is compatible with the existence of a WIMP dark matter candidate charged under $U(1)'$. For simplicity, we focus on regions of parameter space where the two sectors are in equilibrium (however, as seen in the previous subsection, even when this does not occur, a sufficient amount of baryon asymmetry can still be produced via freeze-in production of the $\Bp$), and work with the mass hierarchy $\mBp> m_X > m_{Z'}$. 

In this framework, there are two potential concerns: 
\begin{itemize}
\item The $\Bp\Bp\to X \bar{X}$ process, which proceeds with gauge coupling $g_D$ strength, can keep the $\Bp$ in equilibrium even at temperatures below $\mBp$. This significantly suppresses the $\Bp$ freeze-out abundance $Y_{\Bp}$, which in turn can suppress the baryon asymmetry. 
\item If the $\Bp$ has a long enough lifetime that it dominates the energy density of the Universe, the subsequent entropy dilution from its decays can suppress the relic density of dark matter, spoiling the attractive success of the ``WIMP miracle".
\end{itemize}

\begin{figure}[t]
\includegraphics[width=4in]{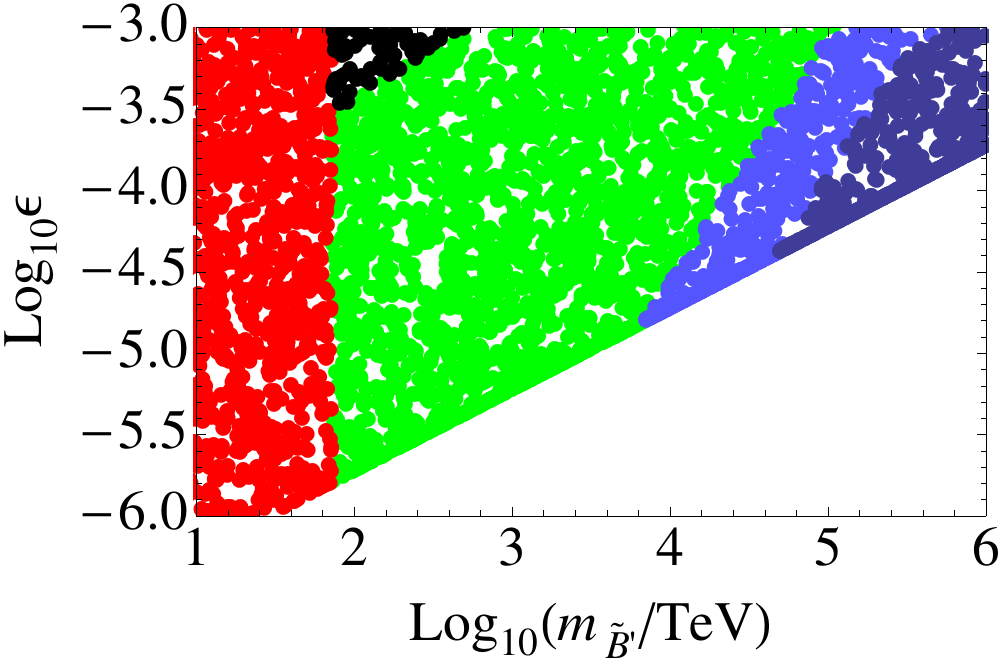}
\caption{\label{fig:fullscanextended} Various regions of parameter space in the extended hidden sector framework with WIMP dark matter, using $m_X=1$ TeV (which fixes $g_D\approx 0.5$), $m_0=7\,\mBp,\,~m_{\bino}=0.25\,\mBp,$ and $\lambda''=0.1$. The various colored regions are: the hidden and visible sectors do not thermalize (white); insufficient amount of baryon asymmetry (red); washout processes (BNV inverse decay and/or annihilations) are faster than Hubble at the time of $\Bp$ decay (black); the $\Bp$ dominates the energy density of the Universe before it decays, resulting in a suppression of the dark matter abundance by an $\mathcal{O}(1)$ factor (light blue) or by over an order of magnitude (dark blue). In the green region, the desired amount of baryon asymmetry can be produced while remaining consistent with all of these constraints.}
\end{figure}

We study these effects with a numerical scan. Fig.\,\ref{fig:fullscanextended} shows the various regions of parameter space in the $\mBp - \epsilon$ plane, with the other parameters fixed as specified in the caption. Here, for concreteness we have assumed $m_{\tilde{X}}=m_0$, i.e. the same sfermion masses in both the hidden and visible sectors. This plot shows that both of the above concerns are realized in different parts of parameter space. In the red region, the suppressed $\Bp$ abundance from freeze-out with the dark matter particle makes it impossible to produce the desired amount of baryon asymmetry. In the opposite extreme (blue region), the $\Bp$ dominates the energy density of the Universe at the time of its decay, diluting the abundance of the baryon asymmetry as well as of dark matter. We show this region in two shades: in the light blue region, the abundance is diluted only by an $\mathcal{O}(1)$ factor, so that the correct dark matter abundance can still be realized via the WIMP miracle; in the dark blue region, abundances are diluted by over an order of magnitude, and the connection with the WIMP miracle is more tenuous. Nevertheless, we see that the baryogenesis mechanism studied in this paper is compatible with hidden sector WIMP dark matter in large regions of parameter space (green region). The shapes of these regions can be understood by noting that, with $m_X$ and $m_0/\mBp$ fixed, $Y_{\Bp}$ is proportional to $\mBp$, see Eq.\,(\ref{eq:yextended}).

\section{Complementary Phenomenology}
\label{sec:others}

In this section, we briefly discuss some other phenomenological signatures that might be possible within the baryogenesis framework discussed in this paper. As we saw in the previous section, the desired amount of baryon asymmetry can be produced over a large range of mass scales. While experimental evidence will prove difficult for heavy scales, additional phenomenology is possible on several fronts if the superpartners are light, as we now describe. 

\vskip 0.2cm
\noindent\textit{Collider and Direct Signatures:}

There are stringent constraints on squarks from the LHC even with RPV  \cite{Dercks:2017lfq,Aaboud:2017dmy,Aaboud:2017nmi,Bardhan:2016gui,Li:2018qxr}, where resonant production provides additional search modes \cite{Monteux:2016gag}. Under various assumptions, such constraints currently lie at the $\mathcal{O}(100)$ GeV - $1$ TeV scale.  In our studies in Sec.\,\ref{sec:results}, we restricted $m_0$ to lie above $1$ TeV in light of such constraints, but baryogenesis can be realized with lower masses. In our framework, TeV scale squarks (assuming a large enough Higgs boson mass can be generated) can therefore be accessible to collider searches, e.g. for $\tilde{q}\to q \chi_0 \to q(qqq)$. This is in contrast with the MSSM implementation of baryogenesis in \cite{Arcadi:2015ffa,Cui:2013bta}, where the sfermions are required to be several orders of magnitude heavier in order to satisfy the out of equilibrium condition and are therefore beyond the reach of colliders. 

Likewise, direct searches also constrain the existence of a kinetically mixed light $Z'$. Recall that the gauge boson kinetic mixing angle $\eta$ is in principle different from the gaugino mixing angle $\epsilon$, but one generally expects $\eta\sim\epsilon$. Current limits (see Fig.6 in \cite{Evans:2017kti}) are at the level of $\eta\sim 10^{-3}$ for GeV scale $Z'$ (deteriorating to $\eta\sim 10^{-2}$ above 100 GeV), but much stronger below the GeV scale due to beam dump and SN1987A limits. In our framework, such probes are therefore relevant only if the $Z'$ is extremely light (below GeV scale), or if $\epsilon\sim 10^{-3}$.

\vskip 0.2cm
\noindent\textit{Low Energy Signatures:}

The presence of light superpartners with RPV couplings can give rise to several low energy signatures. These depend on the flavor structure of the RPV coupling $\lambda_{ijk}$, i.e., which quarks are involved in the RPV interactions. We briefly discuss the most interesting signatures relevant to our framework here, and refer the interested reader to \cite{Barbier:2004ez} for a more comprehensive discussion of various low energy constraints on RPV supersymmetry.

If the coupling involves first generation quarks, a particularly strong test of baryon number violation comes from searches for neutron-antineutron oscillations $n-\bar{n}$.\,\footnote{For studies discussing $n-\bar{n}$ signals in the context of baryogenesis scenarios, see \cite{Calibbi:2017rab,Cheung:2013hza,Grojean:2018fus}.} In our scenario, since the coupling of the $\Bp$ to SM is suppressed by $\epsilon$, the $n-\bar{n}$ oscillation is primarily mediated by the MSSM bino $\bino$, via effective $\bino u d d$ four-fermion couplings obtained by integrating out the squark. Note that the RPV coupling $\lambda''_{111}=0$, hence generating this effective vertex requires a flavor off-diagonal mixing term in the squark sector, leading to a suppression of the signal \cite{Barbier:2004ez}. Current $n-\bar{n}$ constraints \cite{BaldoCeolin:1994jz,Abe:2011ky} can be translated to the following bound on this off-diagonal mixing angle:
\beq
\tilde{\theta}_{1j}\lsim \left(\frac{m_0}{100\,\text{TeV}}\right)^2 \left(\frac{\mBp}{\text{TeV}}\right)^{0.5} \left(\frac{0.01}{\lambda''}\right).
\eeq 
Upcoming experiments \cite{Milstead:2015toa,Frost:2016qzt} will improve on this sensitivity by more than an order of magnitude and could probe our framework in the presence of such flavor off-diagonal mixing.

There are other processes that do not suffer from this unknown mixing suppression, such as double nucleon decays (such as $p\,p\to K^+K^+$ or $n\,n\to K^0\,K^0$). Such processes most strongly constrain $\lambda''_{112},\lambda''_{113}\,(\,\lsim 10^{-4},10^{-1}$ respectively for TeV scale superpartners), but involve large uncertainties from hadronic and nuclear matrix elements, see \cite{Goity:1994dq,Barbier:2004ez} for details. 

 EDM measurements are a powerful probe of CP violation and are known to constrain even PeV scale sfermions \cite{Altmannshofer:2013lfa,Cesarotti:2018huy}. In our framework, the EDM contribution arises through a gaugino-sfermion loop diagram. However, since the CP phase relevant to baryogenesis resides in the combination of the two gaugino masses, Arg$(\mBp m_{\bino}^*)$, both gaugino mass insertions are required on the gaugino propagator in the loop to generate the EDMs. This suppresses the amplitude by $\epsilon^2$, greatly suppressing the EDMs, which can be estimated (for down-type fermions) as:
\beq
\frac{d_{f}}{10^{-29}\,e\,\text{cm}} \approx \text{Im}[e^{2i\theta}]\left(\frac{m_f}{\text{MeV}}\right)\left(\frac{\epsilon}{10^{-3}}\right)^2\left(\frac{\text{TeV}}{m_0}\right)^6
\left(\frac{\mBp}{\text{TeV}}\right)^2\left(\frac{m_{\bino}}{\text{TeV}}\right)\left(\frac{\mu}{100\,\text{TeV}}\right)\text{tan}\,\beta.
\eeq
Future experimental limits are expected to reach $d_{e,p,n}\sim 10^{-29}$\,e\,cm \cite{Hewett:2012ns}, with electron, proton, and neutron EDMs expected to provide comparable reach if all sfermions are at the same scale.  We see that the $\epsilon^2$ contribution can suppress the EDMs below observable limits even for TeV scale sfermions and gauginos\,\footnote{In the presence of off-diagonal flavor mixing in the sfermion sector, the EDMs can get enhanced by various fermion mass ratios \cite{Altmannshofer:2013lfa}.}, making them likely unobservable even with improved future measurements unless $\epsilon\sim10^{-3}$. 

\vskip 0.2cm
\noindent\textit{Dark Matter Signatures:}

If the hidden sector also contains a WIMP dark matter particle, as in the extended hidden sector scenario we considered, additional signatures are possible at both direct and indirect detection experiments. 

The Dirac nature of the dark matter particle $X$ gives rise to a spin-independent direct detection cross section with nuclei mediated by $Z, Z'$ gauge bosons; this cross section scales approximately as (see \cite{Essig:2007az,Evans:2017kti} for details)
\beq
\sigma_{DD, SI}\sim 10^{-39}\, g_D^2\, \epsilon^2\left(\frac{m_Z}{m_{Z'}}\right)^4 \text{cm}^2.
\eeq
This could provide an observable signal if $\epsilon\lsim10^{-3}$ and $m_{Z'}\sim m_Z$ (See e.g. the projected sensitivity in Fig.\,6 of \cite{Evans:2017kti}.). 

Indirect detection signals, on the other hand, hold more promise as they are not suppressed by the small $\epsilon$ parameter. The dark matter annihilation process  $X\bar{X}\to Z' Z'$ proceeds with weak scale cross sections, and the $Z'$ subsequently decay into SM fermions, yielding observable signals at indirect detection instruments. Such cascade decays of hidden sector dark matter along with prospects/constraints from various indirect detection searches, such as with CMB measurements, from dwarf galaxies, at AMS-02, or at Cherenkov telescopes, have been studied extensively in \cite{Elor:2015bho} (see also \cite{Profumo:2017obk,Evans:2017kti,Leane:2018kjk,Cline:2014dwa}). These studies suggest that such measurements can probe TeV scale hidden sector WIMP dark matter.

\section{Summary}
\label{sec:summary}

In this paper, we studied baryogenesis via a gaugino portal, with out-of-equilibrium decays of a hidden sector gaugino $\Bp$ via its portal mixing with the MSSM bino producing the  baryon asymmetry of the Universe.   We summarize our main findings below:

\begin{itemize}
\item This mechanism can produce the desired baryon asymmetry in a minimal hidden sector framework, where all relevant phenomenology is controlled by the portal coupling of the $\Bp$ with the visible sector, as well as within an extended hidden sector that contains additional particles interacting with the $\Bp$. 
In particular, the baryogenesis mechanism can coexist with hidden sector WIMP dark matter, which is an attractive possibility given the lack of a dark matter candidate in the visible sector due to R-parity violation. 

\item The desired baryon asymmetry can be produced across wide ranges of values of $\mBp$ and $\epsilon$, spanning several orders of magnitude. In particular, the mechanism is consistent with $\mBp$ as low as $10$ GeV. Such a low scale baryogenesis mechanism is an attractive prospect due to the possibility of additional phenomenological signatures.

\item The small portal coupling $\epsilon$ is crucial in two respects. It makes the $\Bp$ sufficiently long-lived to evade washout of the baryon asymmetry from inverse decay and annihilation effects without suppressing $\epsilon_{CP}$, the fraction of $\Bp$ decays that produce a baryon asymmetry. Furthermore, while CP violating low scale physics generally gives rise to too-large EDMs, in this framework the EDMs are also $\epsilon^2$ suppressed, making a low scale baryogenesis implementation consistent with stringent EDM constraints.  

\end{itemize}

It is attractive that baryogenesis from long-lived particle decays, previously studied in other contexts, may be naturally implemented in a framework involving supersymmetry and a hidden sector, both well-motivated for other reasons.

Our study contains several aspects  worth pursuing in greater detail: for instance, it will be interesting to explore other hidden sector dark matter candidates that are compatible with baryogenesis discussed here. Our work also contains several possible experimental signatures that might motivate additional study, ranging from colliders, direct searches, low energy probes, to indirect dark matter searches. These are broadly of two kinds. There are those whose observability requires $\epsilon \gsim 10^{-3}$ (close to the upper limit we have studied); these includes signals such as direct $Z'$ production, EDMs, and direct detection of dark matter. However, there are other experimental signatures that are independent of the portal coupling -- squarks at the LHC, neutron antineutron oscillation, and indirect detection of dark matter -- which can therefore be observable even with much smaller values of $\epsilon$.

\section*{Acknowledgements}

We thank Wolfgang Altmannshofer, Stefania Gori, and Raman Sundrum for helpful conversations. The work of A.P. was supported by the DoE under grant DE-SC0007859.  BS is partially supported by the NSF CAREER grant PHY-1654502.

\bibliography{gauginoportal}{}

\begin{thebibliography}{53}%
\makeatletter
\providecommand \@ifxundefined [1]{%
 \@ifx{#1\undefined}
}%
\providecommand \@ifnum [1]{%
 \ifnum #1\expandafter \@firstoftwo
 \else \expandafter \@secondoftwo
 \fi
}%
\providecommand \@ifx [1]{%
 \ifx #1\expandafter \@firstoftwo
 \else \expandafter \@secondoftwo
 \fi
}%
\providecommand \natexlab [1]{#1}%
\providecommand \enquote  [1]{``#1''}%
\providecommand \bibnamefont  [1]{#1}%
\providecommand \bibfnamefont [1]{#1}%
\providecommand \citenamefont [1]{#1}%
\providecommand \href@noop [0]{\@secondoftwo}%
\providecommand \href [0]{\begingroup \@sanitize@url \@href}%
\providecommand \@href[1]{\@@startlink{#1}\@@href}%
\providecommand \@@href[1]{\endgroup#1\@@endlink}%
\providecommand \@sanitize@url [0]{\catcode `\\12\catcode `\$12\catcode
  `\&12\catcode `\#12\catcode `\^12\catcode `\_12\catcode `\%12\relax}%
\providecommand \@@startlink[1]{}%
\providecommand \@@endlink[0]{}%
\providecommand \url  [0]{\begingroup\@sanitize@url \@url }%
\providecommand \@url [1]{\endgroup\@href {#1}{\urlprefix }}%
\providecommand \urlprefix  [0]{URL }%
\providecommand \Eprint [0]{\href }%
\providecommand \doibase [0]{http://dx.doi.org/}%
\providecommand \selectlanguage [0]{\@gobble}%
\providecommand \bibinfo  [0]{\@secondoftwo}%
\providecommand \bibfield  [0]{\@secondoftwo}%
\providecommand \translation [1]{[#1]}%
\providecommand \BibitemOpen [0]{}%
\providecommand \bibitemStop [0]{}%
\providecommand \bibitemNoStop [0]{.\EOS\space}%
\providecommand \EOS [0]{\spacefactor3000\relax}%
\providecommand \BibitemShut  [1]{\csname bibitem#1\endcsname}%
\let\auto@bib@innerbib\@empty
\bibitem [{\citenamefont {Alexander}\ \emph {et~al.}(2016)\citenamefont
  {Alexander} \emph {et~al.}}]{Alexander:2016aln}%
  \BibitemOpen
  \bibfield  {author} {\bibinfo {author} {\bibfnamefont {J.}~\bibnamefont
  {Alexander}} \emph {et~al.}\ }(\bibinfo {year} {2016})\ \Eprint
  {http://arxiv.org/abs/1608.08632} {arXiv:1608.08632 [hep-ph]} \BibitemShut
  {NoStop}%
\bibitem [{\citenamefont {Holdom}(1986)}]{Holdom:1985ag}%
  \BibitemOpen
  \bibfield  {author} {\bibinfo {author} {\bibfnamefont {B.}~\bibnamefont
  {Holdom}},\ }\href {\doibase 10.1016/0370-2693(86)91377-8} {\bibfield
  {journal} {\bibinfo  {journal} {Phys. Lett.}\ }\textbf {\bibinfo {volume}
  {166B}},\ \bibinfo {pages} {196} (\bibinfo {year} {1986})}\BibitemShut
  {NoStop}%
\bibitem [{\citenamefont {Langacker}(2009)}]{Langacker:2008yv}%
  \BibitemOpen
  \bibfield  {author} {\bibinfo {author} {\bibfnamefont {P.}~\bibnamefont
  {Langacker}},\ }\href {\doibase 10.1103/RevModPhys.81.1199} {\bibfield
  {journal} {\bibinfo  {journal} {Rev. Mod. Phys.}\ }\textbf {\bibinfo {volume}
  {81}},\ \bibinfo {pages} {1199} (\bibinfo {year} {2009})},\ \Eprint
  {http://arxiv.org/abs/0801.1345} {arXiv:0801.1345 [hep-ph]} \BibitemShut
  {NoStop}%
\bibitem [{\citenamefont {Dienes}\ \emph {et~al.}(1997)\citenamefont {Dienes},
  \citenamefont {Kolda},\ and\ \citenamefont {March-Russell}}]{Dienes:1996zr}%
  \BibitemOpen
  \bibfield  {author} {\bibinfo {author} {\bibfnamefont {K.~R.}\ \bibnamefont
  {Dienes}}, \bibinfo {author} {\bibfnamefont {C.~F.}\ \bibnamefont {Kolda}}, \
  and\ \bibinfo {author} {\bibfnamefont {J.}~\bibnamefont {March-Russell}},\
  }\href {\doibase 10.1016/S0550-3213(97)80028-4,
  10.1016/S0550-3213(97)00173-9} {\bibfield  {journal} {\bibinfo  {journal}
  {Nucl. Phys.}\ }\textbf {\bibinfo {volume} {B492}},\ \bibinfo {pages} {104}
  (\bibinfo {year} {1997})},\ \Eprint {http://arxiv.org/abs/hep-ph/9610479}
  {arXiv:hep-ph/9610479 [hep-ph]} \BibitemShut {NoStop}%
\bibitem [{\citenamefont {Sakharov}(1967)}]{Sakharov:1967dj}%
  \BibitemOpen
  \bibfield  {author} {\bibinfo {author} {\bibfnamefont {A.~D.}\ \bibnamefont
  {Sakharov}},\ }\href {\doibase 10.1070/PU1991v034n05ABEH002497} {\bibfield
  {journal} {\bibinfo  {journal} {Pisma Zh. Eksp. Teor. Fiz.}\ }\textbf
  {\bibinfo {volume} {5}},\ \bibinfo {pages} {32} (\bibinfo {year} {1967})},\
  \bibinfo {note} {[Usp. Fiz. Nauk161,no.5,61(1991)]}\BibitemShut {NoStop}%
\bibitem [{\citenamefont {Dimopoulos}\ and\ \citenamefont
  {Hall}(1987)}]{Dimopoulos:1987rk}%
  \BibitemOpen
  \bibfield  {author} {\bibinfo {author} {\bibfnamefont {S.}~\bibnamefont
  {Dimopoulos}}\ and\ \bibinfo {author} {\bibfnamefont {L.~J.}\ \bibnamefont
  {Hall}},\ }\href {\doibase 10.1016/0370-2693(87)90593-4} {\bibfield
  {journal} {\bibinfo  {journal} {Phys. Lett.}\ }\textbf {\bibinfo {volume}
  {B196}},\ \bibinfo {pages} {135} (\bibinfo {year} {1987})}\BibitemShut
  {NoStop}%
\bibitem [{\citenamefont {Claudson}\ \emph {et~al.}(1984)\citenamefont
  {Claudson}, \citenamefont {Hall},\ and\ \citenamefont
  {Hinchliffe}}]{Claudson:1983js}%
  \BibitemOpen
  \bibfield  {author} {\bibinfo {author} {\bibfnamefont {M.}~\bibnamefont
  {Claudson}}, \bibinfo {author} {\bibfnamefont {L.~J.}\ \bibnamefont {Hall}},
  \ and\ \bibinfo {author} {\bibfnamefont {I.}~\bibnamefont {Hinchliffe}},\
  }\href {\doibase 10.1016/0550-3213(84)90212-8} {\bibfield  {journal}
  {\bibinfo  {journal} {Nucl. Phys.}\ }\textbf {\bibinfo {volume} {B241}},\
  \bibinfo {pages} {309} (\bibinfo {year} {1984})}\BibitemShut {NoStop}%
\bibitem [{\citenamefont {Rompineve}(2014)}]{Sorbello:2013xwa}%
  \BibitemOpen
  \bibfield  {author} {\bibinfo {author} {\bibfnamefont {F.}~\bibnamefont
  {Rompineve}},\ }\href {\doibase 10.1007/JHEP08(2014)014} {\bibfield
  {journal} {\bibinfo  {journal} {JHEP}\ }\textbf {\bibinfo {volume} {08}},\
  \bibinfo {pages} {014} (\bibinfo {year} {2014})},\ \Eprint
  {http://arxiv.org/abs/1310.0840} {arXiv:1310.0840 [hep-ph]} \BibitemShut
  {NoStop}%
\bibitem [{\citenamefont {Cui}\ and\ \citenamefont
  {Sundrum}(2013)}]{Cui:2012jh}%
  \BibitemOpen
  \bibfield  {author} {\bibinfo {author} {\bibfnamefont {Y.}~\bibnamefont
  {Cui}}\ and\ \bibinfo {author} {\bibfnamefont {R.}~\bibnamefont {Sundrum}},\
  }\href {\doibase 10.1103/PhysRevD.87.116013} {\bibfield  {journal} {\bibinfo
  {journal} {Phys. Rev.}\ }\textbf {\bibinfo {volume} {D87}},\ \bibinfo {pages}
  {116013} (\bibinfo {year} {2013})},\ \Eprint {http://arxiv.org/abs/1212.2973}
  {arXiv:1212.2973 [hep-ph]} \BibitemShut {NoStop}%
\bibitem [{\citenamefont {Cui}(2013)}]{Cui:2013bta}%
  \BibitemOpen
  \bibfield  {author} {\bibinfo {author} {\bibfnamefont {Y.}~\bibnamefont
  {Cui}},\ }\href {\doibase 10.1007/JHEP12(2013)067} {\bibfield  {journal}
  {\bibinfo  {journal} {JHEP}\ }\textbf {\bibinfo {volume} {12}},\ \bibinfo
  {pages} {067} (\bibinfo {year} {2013})},\ \Eprint
  {http://arxiv.org/abs/1309.2952} {arXiv:1309.2952 [hep-ph]} \BibitemShut
  {NoStop}%
\bibitem [{\citenamefont {Arcadi}\ \emph {et~al.}(2015)\citenamefont {Arcadi},
  \citenamefont {Covi},\ and\ \citenamefont {Nardecchia}}]{Arcadi:2015ffa}%
  \BibitemOpen
  \bibfield  {author} {\bibinfo {author} {\bibfnamefont {G.}~\bibnamefont
  {Arcadi}}, \bibinfo {author} {\bibfnamefont {L.}~\bibnamefont {Covi}}, \ and\
  \bibinfo {author} {\bibfnamefont {M.}~\bibnamefont {Nardecchia}},\ }\href
  {\doibase 10.1103/PhysRevD.92.115006} {\bibfield  {journal} {\bibinfo
  {journal} {Phys. Rev.}\ }\textbf {\bibinfo {volume} {D92}},\ \bibinfo {pages}
  {115006} (\bibinfo {year} {2015})},\ \Eprint
  {http://arxiv.org/abs/1507.05584} {arXiv:1507.05584 [hep-ph]} \BibitemShut
  {NoStop}%
\bibitem [{\citenamefont {Barbier}\ \emph {et~al.}(2005)\citenamefont {Barbier}
  \emph {et~al.}}]{Barbier:2004ez}%
  \BibitemOpen
  \bibfield  {author} {\bibinfo {author} {\bibfnamefont {R.}~\bibnamefont
  {Barbier}} \emph {et~al.},\ }\href {\doibase 10.1016/j.physrep.2005.08.006}
  {\bibfield  {journal} {\bibinfo  {journal} {Phys. Rept.}\ }\textbf {\bibinfo
  {volume} {420}},\ \bibinfo {pages} {1} (\bibinfo {year} {2005})},\ \Eprint
  {http://arxiv.org/abs/hep-ph/0406039} {arXiv:hep-ph/0406039 [hep-ph]}
  \BibitemShut {NoStop}%
\bibitem [{\citenamefont {Altmannshofer}\ \emph {et~al.}(2013)\citenamefont
  {Altmannshofer}, \citenamefont {Harnik},\ and\ \citenamefont
  {Zupan}}]{Altmannshofer:2013lfa}%
  \BibitemOpen
  \bibfield  {author} {\bibinfo {author} {\bibfnamefont {W.}~\bibnamefont
  {Altmannshofer}}, \bibinfo {author} {\bibfnamefont {R.}~\bibnamefont
  {Harnik}}, \ and\ \bibinfo {author} {\bibfnamefont {J.}~\bibnamefont
  {Zupan}},\ }\href {\doibase 10.1007/JHEP11(2013)202} {\bibfield  {journal}
  {\bibinfo  {journal} {JHEP}\ }\textbf {\bibinfo {volume} {11}},\ \bibinfo
  {pages} {202} (\bibinfo {year} {2013})},\ \Eprint
  {http://arxiv.org/abs/1308.3653} {arXiv:1308.3653 [hep-ph]} \BibitemShut
  {NoStop}%
\bibitem [{\citenamefont {Cesarotti}\ \emph {et~al.}(2018)\citenamefont
  {Cesarotti}, \citenamefont {Lu}, \citenamefont {Nakai}, \citenamefont
  {Parikh},\ and\ \citenamefont {Reece}}]{Cesarotti:2018huy}%
  \BibitemOpen
  \bibfield  {author} {\bibinfo {author} {\bibfnamefont {C.}~\bibnamefont
  {Cesarotti}}, \bibinfo {author} {\bibfnamefont {Q.}~\bibnamefont {Lu}},
  \bibinfo {author} {\bibfnamefont {Y.}~\bibnamefont {Nakai}}, \bibinfo
  {author} {\bibfnamefont {A.}~\bibnamefont {Parikh}}, \ and\ \bibinfo {author}
  {\bibfnamefont {M.}~\bibnamefont {Reece}},\ }\href@noop {} {\  (\bibinfo
  {year} {2018})},\ \Eprint {http://arxiv.org/abs/1810.07736} {arXiv:1810.07736
  [hep-ph]} \BibitemShut {NoStop}%
\bibitem [{\citenamefont {Ibarra}\ \emph {et~al.}(2009)\citenamefont {Ibarra},
  \citenamefont {Ringwald},\ and\ \citenamefont {Weniger}}]{Ibarra:2008kn}%
  \BibitemOpen
  \bibfield  {author} {\bibinfo {author} {\bibfnamefont {A.}~\bibnamefont
  {Ibarra}}, \bibinfo {author} {\bibfnamefont {A.}~\bibnamefont {Ringwald}}, \
  and\ \bibinfo {author} {\bibfnamefont {C.}~\bibnamefont {Weniger}},\ }\href
  {\doibase 10.1088/1475-7516/2009/01/003} {\bibfield  {journal} {\bibinfo
  {journal} {JCAP}\ }\textbf {\bibinfo {volume} {0901}},\ \bibinfo {pages}
  {003} (\bibinfo {year} {2009})},\ \Eprint {http://arxiv.org/abs/0809.3196}
  {arXiv:0809.3196 [hep-ph]} \BibitemShut {NoStop}%
\bibitem [{\citenamefont {Arvanitaki}\ \emph {et~al.}(2010)\citenamefont
  {Arvanitaki}, \citenamefont {Craig}, \citenamefont {Dimopoulos},
  \citenamefont {Dubovsky},\ and\ \citenamefont
  {March-Russell}}]{Arvanitaki:2009hb}%
  \BibitemOpen
  \bibfield  {author} {\bibinfo {author} {\bibfnamefont {A.}~\bibnamefont
  {Arvanitaki}}, \bibinfo {author} {\bibfnamefont {N.}~\bibnamefont {Craig}},
  \bibinfo {author} {\bibfnamefont {S.}~\bibnamefont {Dimopoulos}}, \bibinfo
  {author} {\bibfnamefont {S.}~\bibnamefont {Dubovsky}}, \ and\ \bibinfo
  {author} {\bibfnamefont {J.}~\bibnamefont {March-Russell}},\ }\href {\doibase
  10.1103/PhysRevD.81.075018} {\bibfield  {journal} {\bibinfo  {journal} {Phys.
  Rev.}\ }\textbf {\bibinfo {volume} {D81}},\ \bibinfo {pages} {075018}
  (\bibinfo {year} {2010})},\ \Eprint {http://arxiv.org/abs/0909.5440}
  {arXiv:0909.5440 [hep-ph]} \BibitemShut {NoStop}%
\bibitem [{\citenamefont {Acharya}\ \emph {et~al.}(2016)\citenamefont
  {Acharya}, \citenamefont {Ellis}, \citenamefont {Kane}, \citenamefont
  {Nelson},\ and\ \citenamefont {Perry}}]{Acharya:2016fge}%
  \BibitemOpen
  \bibfield  {author} {\bibinfo {author} {\bibfnamefont {B.~S.}\ \bibnamefont
  {Acharya}}, \bibinfo {author} {\bibfnamefont {S.~A.~R.}\ \bibnamefont
  {Ellis}}, \bibinfo {author} {\bibfnamefont {G.~L.}\ \bibnamefont {Kane}},
  \bibinfo {author} {\bibfnamefont {B.~D.}\ \bibnamefont {Nelson}}, \ and\
  \bibinfo {author} {\bibfnamefont {M.~J.}\ \bibnamefont {Perry}},\ }\href
  {\doibase 10.1103/PhysRevLett.117.181802} {\bibfield  {journal} {\bibinfo
  {journal} {Phys. Rev. Lett.}\ }\textbf {\bibinfo {volume} {117}},\ \bibinfo
  {pages} {181802} (\bibinfo {year} {2016})},\ \Eprint
  {http://arxiv.org/abs/1604.05320} {arXiv:1604.05320 [hep-ph]} \BibitemShut
  {NoStop}%
\bibitem [{\citenamefont {Blumenhagen}\ \emph {et~al.}(2006)\citenamefont
  {Blumenhagen}, \citenamefont {Moster},\ and\ \citenamefont
  {Weigand}}]{Blumenhagen:2006ux}%
  \BibitemOpen
  \bibfield  {author} {\bibinfo {author} {\bibfnamefont {R.}~\bibnamefont
  {Blumenhagen}}, \bibinfo {author} {\bibfnamefont {S.}~\bibnamefont {Moster}},
  \ and\ \bibinfo {author} {\bibfnamefont {T.}~\bibnamefont {Weigand}},\ }\href
  {\doibase 10.1016/j.nuclphysb.2006.06.005} {\bibfield  {journal} {\bibinfo
  {journal} {Nucl. Phys.}\ }\textbf {\bibinfo {volume} {B751}},\ \bibinfo
  {pages} {186} (\bibinfo {year} {2006})},\ \Eprint
  {http://arxiv.org/abs/hep-th/0603015} {arXiv:hep-th/0603015 [hep-th]}
  \BibitemShut {NoStop}%
\bibitem [{\citenamefont {Lust}\ and\ \citenamefont
  {Stieberger}(2007)}]{Lust:2003ky}%
  \BibitemOpen
  \bibfield  {author} {\bibinfo {author} {\bibfnamefont {D.}~\bibnamefont
  {Lust}}\ and\ \bibinfo {author} {\bibfnamefont {S.}~\bibnamefont
  {Stieberger}},\ }\href {\doibase 10.1002/prop.200310335} {\bibfield
  {journal} {\bibinfo  {journal} {Fortsch. Phys.}\ }\textbf {\bibinfo {volume}
  {55}},\ \bibinfo {pages} {427} (\bibinfo {year} {2007})},\ \Eprint
  {http://arxiv.org/abs/hep-th/0302221} {arXiv:hep-th/0302221 [hep-th]}
  \BibitemShut {NoStop}%
\bibitem [{\citenamefont {Abel}\ and\ \citenamefont
  {Schofield}(2004)}]{Abel:2003ue}%
  \BibitemOpen
  \bibfield  {author} {\bibinfo {author} {\bibfnamefont {S.~A.}\ \bibnamefont
  {Abel}}\ and\ \bibinfo {author} {\bibfnamefont {B.~W.}\ \bibnamefont
  {Schofield}},\ }\href {\doibase 10.1016/j.nuclphysb.2004.02.037} {\bibfield
  {journal} {\bibinfo  {journal} {Nucl. Phys.}\ }\textbf {\bibinfo {volume}
  {B685}},\ \bibinfo {pages} {150} (\bibinfo {year} {2004})},\ \Eprint
  {http://arxiv.org/abs/hep-th/0311051} {arXiv:hep-th/0311051 [hep-th]}
  \BibitemShut {NoStop}%
\bibitem [{\citenamefont {Berg}\ \emph {et~al.}(2005)\citenamefont {Berg},
  \citenamefont {Haack},\ and\ \citenamefont {Kors}}]{Berg:2004ek}%
  \BibitemOpen
  \bibfield  {author} {\bibinfo {author} {\bibfnamefont {M.}~\bibnamefont
  {Berg}}, \bibinfo {author} {\bibfnamefont {M.}~\bibnamefont {Haack}}, \ and\
  \bibinfo {author} {\bibfnamefont {B.}~\bibnamefont {Kors}},\ }\href {\doibase
  10.1103/PhysRevD.71.026005} {\bibfield  {journal} {\bibinfo  {journal} {Phys.
  Rev.}\ }\textbf {\bibinfo {volume} {D71}},\ \bibinfo {pages} {026005}
  (\bibinfo {year} {2005})},\ \Eprint {http://arxiv.org/abs/hep-th/0404087}
  {arXiv:hep-th/0404087 [hep-th]} \BibitemShut {NoStop}%
\bibitem [{\citenamefont {Abel}\ \emph
  {et~al.}(2008{\natexlab{a}})\citenamefont {Abel}, \citenamefont {Jaeckel},
  \citenamefont {Khoze},\ and\ \citenamefont {Ringwald}}]{Abel:2006qt}%
  \BibitemOpen
  \bibfield  {author} {\bibinfo {author} {\bibfnamefont {S.~A.}\ \bibnamefont
  {Abel}}, \bibinfo {author} {\bibfnamefont {J.}~\bibnamefont {Jaeckel}},
  \bibinfo {author} {\bibfnamefont {V.~V.}\ \bibnamefont {Khoze}}, \ and\
  \bibinfo {author} {\bibfnamefont {A.}~\bibnamefont {Ringwald}},\ }\href
  {\doibase 10.1016/j.physletb.2008.03.076} {\bibfield  {journal} {\bibinfo
  {journal} {Phys. Lett.}\ }\textbf {\bibinfo {volume} {B666}},\ \bibinfo
  {pages} {66} (\bibinfo {year} {2008}{\natexlab{a}})},\ \Eprint
  {http://arxiv.org/abs/hep-ph/0608248} {arXiv:hep-ph/0608248 [hep-ph]}
  \BibitemShut {NoStop}%
\bibitem [{\citenamefont {Abel}\ \emph
  {et~al.}(2008{\natexlab{b}})\citenamefont {Abel}, \citenamefont {Goodsell},
  \citenamefont {Jaeckel}, \citenamefont {Khoze},\ and\ \citenamefont
  {Ringwald}}]{Abel:2008ai}%
  \BibitemOpen
  \bibfield  {author} {\bibinfo {author} {\bibfnamefont {S.~A.}\ \bibnamefont
  {Abel}}, \bibinfo {author} {\bibfnamefont {M.~D.}\ \bibnamefont {Goodsell}},
  \bibinfo {author} {\bibfnamefont {J.}~\bibnamefont {Jaeckel}}, \bibinfo
  {author} {\bibfnamefont {V.~V.}\ \bibnamefont {Khoze}}, \ and\ \bibinfo
  {author} {\bibfnamefont {A.}~\bibnamefont {Ringwald}},\ }\href {\doibase
  10.1088/1126-6708/2008/07/124} {\bibfield  {journal} {\bibinfo  {journal}
  {JHEP}\ }\textbf {\bibinfo {volume} {07}},\ \bibinfo {pages} {124} (\bibinfo
  {year} {2008}{\natexlab{b}})},\ \Eprint {http://arxiv.org/abs/0803.1449}
  {arXiv:0803.1449 [hep-ph]} \BibitemShut {NoStop}%
\bibitem [{\citenamefont {Nanopoulos}\ and\ \citenamefont
  {Weinberg}(1979)}]{Nanopoulos:1979gx}%
  \BibitemOpen
  \bibfield  {author} {\bibinfo {author} {\bibfnamefont {D.~V.}\ \bibnamefont
  {Nanopoulos}}\ and\ \bibinfo {author} {\bibfnamefont {S.}~\bibnamefont
  {Weinberg}},\ }\href {\doibase 10.1103/PhysRevD.20.2484} {\bibfield
  {journal} {\bibinfo  {journal} {Phys. Rev.}\ }\textbf {\bibinfo {volume}
  {D20}},\ \bibinfo {pages} {2484} (\bibinfo {year} {1979})}\BibitemShut
  {NoStop}%
\bibitem [{\citenamefont {Baldes}\ \emph {et~al.}(2014)\citenamefont {Baldes},
  \citenamefont {Bell}, \citenamefont {Millar}, \citenamefont {Petraki},\ and\
  \citenamefont {Volkas}}]{Baldes:2014rda}%
  \BibitemOpen
  \bibfield  {author} {\bibinfo {author} {\bibfnamefont {I.}~\bibnamefont
  {Baldes}}, \bibinfo {author} {\bibfnamefont {N.~F.}\ \bibnamefont {Bell}},
  \bibinfo {author} {\bibfnamefont {A.}~\bibnamefont {Millar}}, \bibinfo
  {author} {\bibfnamefont {K.}~\bibnamefont {Petraki}}, \ and\ \bibinfo
  {author} {\bibfnamefont {R.~R.}\ \bibnamefont {Volkas}},\ }\href {\doibase
  10.1088/1475-7516/2014/11/041} {\bibfield  {journal} {\bibinfo  {journal}
  {JCAP}\ }\textbf {\bibinfo {volume} {1411}},\ \bibinfo {pages} {041}
  (\bibinfo {year} {2014})},\ \Eprint {http://arxiv.org/abs/1410.0108}
  {arXiv:1410.0108 [hep-ph]} \BibitemShut {NoStop}%
\bibitem [{\citenamefont {Wells}(2003)}]{Wells:2003tf}%
  \BibitemOpen
  \bibfield  {author} {\bibinfo {author} {\bibfnamefont {J.~D.}\ \bibnamefont
  {Wells}},\ }\href@noop {} {\  (\bibinfo {year} {2003})},\ \Eprint
  {http://arxiv.org/abs/hep-ph/0306127} {arXiv:hep-ph/0306127 [hep-ph]}
  \BibitemShut {NoStop}%
\bibitem [{\citenamefont {Arkani-Hamed}\ and\ \citenamefont
  {Dimopoulos}(2005)}]{ArkaniHamed:2004fb}%
  \BibitemOpen
  \bibfield  {author} {\bibinfo {author} {\bibfnamefont {N.}~\bibnamefont
  {Arkani-Hamed}}\ and\ \bibinfo {author} {\bibfnamefont {S.}~\bibnamefont
  {Dimopoulos}},\ }\href {\doibase 10.1088/1126-6708/2005/06/073} {\bibfield
  {journal} {\bibinfo  {journal} {JHEP}\ }\textbf {\bibinfo {volume} {0506}},\
  \bibinfo {pages} {073} (\bibinfo {year} {2005})},\ \Eprint
  {http://arxiv.org/abs/hep-th/0405159} {arXiv:hep-th/0405159 [hep-th]}
  \BibitemShut {NoStop}%
\bibitem [{\citenamefont {Giudice}\ and\ \citenamefont
  {Romanino}(2004)}]{Giudice:2004tc}%
  \BibitemOpen
  \bibfield  {author} {\bibinfo {author} {\bibfnamefont {G.}~\bibnamefont
  {Giudice}}\ and\ \bibinfo {author} {\bibfnamefont {A.}~\bibnamefont
  {Romanino}},\ }\href {\doibase 10.1016/j.nuclphysb.2004.11.048} {\bibfield
  {journal} {\bibinfo  {journal} {Nucl.Phys.}\ }\textbf {\bibinfo {volume}
  {B699}},\ \bibinfo {pages} {65} (\bibinfo {year} {2004})},\ \Eprint
  {http://arxiv.org/abs/hep-ph/0406088} {arXiv:hep-ph/0406088 [hep-ph]}
  \BibitemShut {NoStop}%
\bibitem [{\citenamefont {Wells}(2005)}]{Wells:2004di}%
  \BibitemOpen
  \bibfield  {author} {\bibinfo {author} {\bibfnamefont {J.~D.}\ \bibnamefont
  {Wells}},\ }\href {\doibase 10.1103/PhysRevD.71.015013} {\bibfield  {journal}
  {\bibinfo  {journal} {Phys.Rev.}\ }\textbf {\bibinfo {volume} {D71}},\
  \bibinfo {pages} {015013} (\bibinfo {year} {2005})},\ \Eprint
  {http://arxiv.org/abs/hep-ph/0411041} {arXiv:hep-ph/0411041 [hep-ph]}
  \BibitemShut {NoStop}%
\bibitem [{\citenamefont {Hall}\ \emph {et~al.}(2010)\citenamefont {Hall},
  \citenamefont {Jedamzik}, \citenamefont {March-Russell},\ and\ \citenamefont
  {West}}]{Hall:2009bx}%
  \BibitemOpen
  \bibfield  {author} {\bibinfo {author} {\bibfnamefont {L.~J.}\ \bibnamefont
  {Hall}}, \bibinfo {author} {\bibfnamefont {K.}~\bibnamefont {Jedamzik}},
  \bibinfo {author} {\bibfnamefont {J.}~\bibnamefont {March-Russell}}, \ and\
  \bibinfo {author} {\bibfnamefont {S.~M.}\ \bibnamefont {West}},\ }\href
  {\doibase 10.1007/JHEP03(2010)080} {\bibfield  {journal} {\bibinfo  {journal}
  {JHEP}\ }\textbf {\bibinfo {volume} {03}},\ \bibinfo {pages} {080} (\bibinfo
  {year} {2010})},\ \Eprint {http://arxiv.org/abs/0911.1120} {arXiv:0911.1120
  [hep-ph]} \BibitemShut {NoStop}%
\bibitem [{\citenamefont {Arkani-Hamed}\ \emph {et~al.}(2006)\citenamefont
  {Arkani-Hamed}, \citenamefont {Delgado},\ and\ \citenamefont
  {Giudice}}]{ArkaniHamed:2006mb}%
  \BibitemOpen
  \bibfield  {author} {\bibinfo {author} {\bibfnamefont {N.}~\bibnamefont
  {Arkani-Hamed}}, \bibinfo {author} {\bibfnamefont {A.}~\bibnamefont
  {Delgado}}, \ and\ \bibinfo {author} {\bibfnamefont {G.~F.}\ \bibnamefont
  {Giudice}},\ }\href {\doibase 10.1016/j.nuclphysb.2006.02.010} {\bibfield
  {journal} {\bibinfo  {journal} {Nucl. Phys.}\ }\textbf {\bibinfo {volume}
  {B741}},\ \bibinfo {pages} {108} (\bibinfo {year} {2006})},\ \Eprint
  {http://arxiv.org/abs/hep-ph/0601041} {arXiv:hep-ph/0601041 [hep-ph]}
  \BibitemShut {NoStop}%
\bibitem [{\citenamefont {Kolb}\ and\ \citenamefont
  {Turner}(1990)}]{Kolb:1990vq}%
  \BibitemOpen
  \bibfield  {author} {\bibinfo {author} {\bibfnamefont {E.~W.}\ \bibnamefont
  {Kolb}}\ and\ \bibinfo {author} {\bibfnamefont {M.~S.}\ \bibnamefont
  {Turner}},\ }\href@noop {} {\bibfield  {journal} {\bibinfo  {journal} {Front.
  Phys.}\ }\textbf {\bibinfo {volume} {69}},\ \bibinfo {pages} {1} (\bibinfo
  {year} {1990})}\BibitemShut {NoStop}%
\bibitem [{\citenamefont {Aaboud}\ \emph {et~al.}(2017)\citenamefont {Aaboud}
  \emph {et~al.}}]{Aaboud:2017dmy}%
  \BibitemOpen
  \bibfield  {author} {\bibinfo {author} {\bibfnamefont {M.}~\bibnamefont
  {Aaboud}} \emph {et~al.} (\bibinfo {collaboration} {ATLAS}),\ }\href
  {\doibase 10.1007/JHEP09(2017)084} {\bibfield  {journal} {\bibinfo  {journal}
  {JHEP}\ }\textbf {\bibinfo {volume} {09}},\ \bibinfo {pages} {084} (\bibinfo
  {year} {2017})},\ \Eprint {http://arxiv.org/abs/1706.03731} {arXiv:1706.03731
  [hep-ex]} \BibitemShut {NoStop}%
\bibitem [{\citenamefont {Bardhan}\ \emph {et~al.}(2017)\citenamefont
  {Bardhan}, \citenamefont {Chakraborty}, \citenamefont {Choudhury},
  \citenamefont {Ghosh},\ and\ \citenamefont {Maity}}]{Bardhan:2016gui}%
  \BibitemOpen
  \bibfield  {author} {\bibinfo {author} {\bibfnamefont {D.}~\bibnamefont
  {Bardhan}}, \bibinfo {author} {\bibfnamefont {A.}~\bibnamefont
  {Chakraborty}}, \bibinfo {author} {\bibfnamefont {D.}~\bibnamefont
  {Choudhury}}, \bibinfo {author} {\bibfnamefont {D.~K.}\ \bibnamefont
  {Ghosh}}, \ and\ \bibinfo {author} {\bibfnamefont {M.}~\bibnamefont
  {Maity}},\ }\href {\doibase 10.1103/PhysRevD.96.035024} {\bibfield  {journal}
  {\bibinfo  {journal} {Phys. Rev.}\ }\textbf {\bibinfo {volume} {D96}},\
  \bibinfo {pages} {035024} (\bibinfo {year} {2017})},\ \Eprint
  {http://arxiv.org/abs/1611.03846} {arXiv:1611.03846 [hep-ph]} \BibitemShut
  {NoStop}%
\bibitem [{\citenamefont {Li}\ \emph {et~al.}(2018)\citenamefont {Li},
  \citenamefont {Li},\ and\ \citenamefont {Zhang}}]{Li:2018qxr}%
  \BibitemOpen
  \bibfield  {author} {\bibinfo {author} {\bibfnamefont {J.}~\bibnamefont
  {Li}}, \bibinfo {author} {\bibfnamefont {T.}~\bibnamefont {Li}}, \ and\
  \bibinfo {author} {\bibfnamefont {W.}~\bibnamefont {Zhang}},\ }\href@noop {}
  {\  (\bibinfo {year} {2018})},\ \Eprint {http://arxiv.org/abs/1805.06172}
  {arXiv:1805.06172 [hep-ph]} \BibitemShut {NoStop}%
\bibitem [{\citenamefont {Dercks}\ \emph {et~al.}(2017)\citenamefont {Dercks},
  \citenamefont {Dreiner}, \citenamefont {Krauss}, \citenamefont {Opferkuch},\
  and\ \citenamefont {Reinert}}]{Dercks:2017lfq}%
  \BibitemOpen
  \bibfield  {author} {\bibinfo {author} {\bibfnamefont {D.}~\bibnamefont
  {Dercks}}, \bibinfo {author} {\bibfnamefont {H.}~\bibnamefont {Dreiner}},
  \bibinfo {author} {\bibfnamefont {M.~E.}\ \bibnamefont {Krauss}}, \bibinfo
  {author} {\bibfnamefont {T.}~\bibnamefont {Opferkuch}}, \ and\ \bibinfo
  {author} {\bibfnamefont {A.}~\bibnamefont {Reinert}},\ }\href {\doibase
  10.1140/epjc/s10052-017-5414-4} {\bibfield  {journal} {\bibinfo  {journal}
  {Eur. Phys. J.}\ }\textbf {\bibinfo {volume} {C77}},\ \bibinfo {pages} {856}
  (\bibinfo {year} {2017})},\ \Eprint {http://arxiv.org/abs/1706.09418}
  {arXiv:1706.09418 [hep-ph]} \BibitemShut {NoStop}%
\bibitem [{\citenamefont {Aaboud}\ \emph {et~al.}(2018)\citenamefont {Aaboud}
  \emph {et~al.}}]{Aaboud:2017nmi}%
  \BibitemOpen
  \bibfield  {author} {\bibinfo {author} {\bibfnamefont {M.}~\bibnamefont
  {Aaboud}} \emph {et~al.} (\bibinfo {collaboration} {ATLAS}),\ }\href
  {\doibase 10.1140/epjc/s10052-018-5693-4} {\bibfield  {journal} {\bibinfo
  {journal} {Eur. Phys. J.}\ }\textbf {\bibinfo {volume} {C78}},\ \bibinfo
  {pages} {250} (\bibinfo {year} {2018})},\ \Eprint
  {http://arxiv.org/abs/1710.07171} {arXiv:1710.07171 [hep-ex]} \BibitemShut
  {NoStop}%
\bibitem [{\citenamefont {Monteux}(2016)}]{Monteux:2016gag}%
  \BibitemOpen
  \bibfield  {author} {\bibinfo {author} {\bibfnamefont {A.}~\bibnamefont
  {Monteux}},\ }\href {\doibase 10.1007/JHEP03(2016)216} {\bibfield  {journal}
  {\bibinfo  {journal} {JHEP}\ }\textbf {\bibinfo {volume} {03}},\ \bibinfo
  {pages} {216} (\bibinfo {year} {2016})},\ \Eprint
  {http://arxiv.org/abs/1601.03737} {arXiv:1601.03737 [hep-ph]} \BibitemShut
  {NoStop}%
\bibitem [{\citenamefont {Evans}\ \emph {et~al.}(2018)\citenamefont {Evans},
  \citenamefont {Gori},\ and\ \citenamefont {Shelton}}]{Evans:2017kti}%
  \BibitemOpen
  \bibfield  {author} {\bibinfo {author} {\bibfnamefont {J.~A.}\ \bibnamefont
  {Evans}}, \bibinfo {author} {\bibfnamefont {S.}~\bibnamefont {Gori}}, \ and\
  \bibinfo {author} {\bibfnamefont {J.}~\bibnamefont {Shelton}},\ }\href
  {\doibase 10.1007/JHEP02(2018)100} {\bibfield  {journal} {\bibinfo  {journal}
  {JHEP}\ }\textbf {\bibinfo {volume} {02}},\ \bibinfo {pages} {100} (\bibinfo
  {year} {2018})},\ \Eprint {http://arxiv.org/abs/1712.03974} {arXiv:1712.03974
  [hep-ph]} \BibitemShut {NoStop}%
\bibitem [{\citenamefont {Calibbi}\ \emph {et~al.}(2017)\citenamefont
  {Calibbi}, \citenamefont {Chun},\ and\ \citenamefont
  {Shin}}]{Calibbi:2017rab}%
  \BibitemOpen
  \bibfield  {author} {\bibinfo {author} {\bibfnamefont {L.}~\bibnamefont
  {Calibbi}}, \bibinfo {author} {\bibfnamefont {E.~J.}\ \bibnamefont {Chun}}, \
  and\ \bibinfo {author} {\bibfnamefont {C.~S.}\ \bibnamefont {Shin}},\ }\href
  {\doibase 10.1007/JHEP10(2017)177} {\bibfield  {journal} {\bibinfo  {journal}
  {JHEP}\ }\textbf {\bibinfo {volume} {10}},\ \bibinfo {pages} {177} (\bibinfo
  {year} {2017})},\ \Eprint {http://arxiv.org/abs/1708.06439} {arXiv:1708.06439
  [hep-ph]} \BibitemShut {NoStop}%
\bibitem [{\citenamefont {Cheung}\ and\ \citenamefont
  {Ishiwata}(2013)}]{Cheung:2013hza}%
  \BibitemOpen
  \bibfield  {author} {\bibinfo {author} {\bibfnamefont {C.}~\bibnamefont
  {Cheung}}\ and\ \bibinfo {author} {\bibfnamefont {K.}~\bibnamefont
  {Ishiwata}},\ }\href {\doibase 10.1103/PhysRevD.88.017901} {\bibfield
  {journal} {\bibinfo  {journal} {Phys. Rev.}\ }\textbf {\bibinfo {volume}
  {D88}},\ \bibinfo {pages} {017901} (\bibinfo {year} {2013})},\ \Eprint
  {http://arxiv.org/abs/1304.0468} {arXiv:1304.0468 [hep-ph]} \BibitemShut
  {NoStop}%
\bibitem [{\citenamefont {Grojean}\ \emph {et~al.}(2018)\citenamefont
  {Grojean}, \citenamefont {Shakya}, \citenamefont {Wells},\ and\ \citenamefont
  {Zhang}}]{Grojean:2018fus}%
  \BibitemOpen
  \bibfield  {author} {\bibinfo {author} {\bibfnamefont {C.}~\bibnamefont
  {Grojean}}, \bibinfo {author} {\bibfnamefont {B.}~\bibnamefont {Shakya}},
  \bibinfo {author} {\bibfnamefont {J.~D.}\ \bibnamefont {Wells}}, \ and\
  \bibinfo {author} {\bibfnamefont {Z.}~\bibnamefont {Zhang}},\ }\href@noop {}
  {\  (\bibinfo {year} {2018})},\ \Eprint {http://arxiv.org/abs/1806.00011}
  {arXiv:1806.00011 [hep-ph]} \BibitemShut {NoStop}%
\bibitem [{\citenamefont {Baldo-Ceolin}\ \emph {et~al.}(1994)\citenamefont
  {Baldo-Ceolin} \emph {et~al.}}]{BaldoCeolin:1994jz}%
  \BibitemOpen
  \bibfield  {author} {\bibinfo {author} {\bibfnamefont {M.}~\bibnamefont
  {Baldo-Ceolin}} \emph {et~al.},\ }\href {\doibase 10.1007/BF01580321}
  {\bibfield  {journal} {\bibinfo  {journal} {Z. Phys.}\ }\textbf {\bibinfo
  {volume} {C63}},\ \bibinfo {pages} {409} (\bibinfo {year}
  {1994})}\BibitemShut {NoStop}%
\bibitem [{\citenamefont {Abe}\ \emph {et~al.}(2015)\citenamefont {Abe} \emph
  {et~al.}}]{Abe:2011ky}%
  \BibitemOpen
  \bibfield  {author} {\bibinfo {author} {\bibfnamefont {K.}~\bibnamefont
  {Abe}} \emph {et~al.} (\bibinfo {collaboration} {Super-Kamiokande}),\ }\href
  {\doibase 10.1103/PhysRevD.91.072006} {\bibfield  {journal} {\bibinfo
  {journal} {Phys. Rev.}\ }\textbf {\bibinfo {volume} {D91}},\ \bibinfo {pages}
  {072006} (\bibinfo {year} {2015})},\ \Eprint {http://arxiv.org/abs/1109.4227}
  {arXiv:1109.4227 [hep-ex]} \BibitemShut {NoStop}%
\bibitem [{\citenamefont {Milstead}(2015)}]{Milstead:2015toa}%
  \BibitemOpen
  \bibfield  {author} {\bibinfo {author} {\bibfnamefont {D.}~\bibnamefont
  {Milstead}},\ }\bibfield  {booktitle} {\emph {\bibinfo {booktitle}
  {{Proceedings, 2015 European Physical Society Conference on High Energy
  Physics (EPS-HEP 2015): Vienna, Austria, July 22-29, 2015}}},\ }\href
  {\doibase 10.22323/1.234.0603} {\bibfield  {journal} {\bibinfo  {journal}
  {PoS}\ }\textbf {\bibinfo {volume} {EPS-HEP2015}},\ \bibinfo {pages} {603}
  (\bibinfo {year} {2015})},\ \Eprint {http://arxiv.org/abs/1510.01569}
  {arXiv:1510.01569 [physics.ins-det]} \BibitemShut {NoStop}%
\bibitem [{\citenamefont {Frost}(2017)}]{Frost:2016qzt}%
  \BibitemOpen
  \bibfield  {author} {\bibinfo {author} {\bibfnamefont {M.~J.}\ \bibnamefont
  {Frost}} (\bibinfo {collaboration} {NNbar}),\ }in\ \href {\doibase
  10.1142/9789813148505_0070} {\emph {\bibinfo {booktitle} {{Proceedings, 7th
  Meeting on CPT and Lorentz Symmetry (CPT 16): Bloomington, Indiana, USA, June
  20-24, 2016}}}}\ (\bibinfo {year} {2017})\ pp.\ \bibinfo {pages} {265--267},\
  \Eprint {http://arxiv.org/abs/1607.07271} {arXiv:1607.07271 [hep-ph]}
  \BibitemShut {NoStop}%
\bibitem [{\citenamefont {Goity}\ and\ \citenamefont
  {Sher}(1995)}]{Goity:1994dq}%
  \BibitemOpen
  \bibfield  {author} {\bibinfo {author} {\bibfnamefont {J.~L.}\ \bibnamefont
  {Goity}}\ and\ \bibinfo {author} {\bibfnamefont {M.}~\bibnamefont {Sher}},\
  }\href {\doibase 10.1016/0370-2693(96)01076-3, 10.1016/0370-2693(94)01688-9}
  {\bibfield  {journal} {\bibinfo  {journal} {Phys. Lett.}\ }\textbf {\bibinfo
  {volume} {B346}},\ \bibinfo {pages} {69} (\bibinfo {year} {1995})},\ \bibinfo
  {note} {[Erratum: Phys. Lett.B385,500(1996)]},\ \Eprint
  {http://arxiv.org/abs/hep-ph/9412208} {arXiv:hep-ph/9412208 [hep-ph]}
  \BibitemShut {NoStop}%
\bibitem [{Hew(2012)}]{Hewett:2012ns}%
  \BibitemOpen
  \href {\doibase 10.2172/1042577} {\emph {\bibinfo {title} {{Fundamental
  Physics at the Intensity Frontier}}}}\ (\bibinfo {year} {2012})\ \Eprint
  {http://arxiv.org/abs/1205.2671} {arXiv:1205.2671 [hep-ex]} \BibitemShut
  {NoStop}%
\bibitem [{\citenamefont {Essig}(2008)}]{Essig:2007az}%
  \BibitemOpen
  \bibfield  {author} {\bibinfo {author} {\bibfnamefont {R.}~\bibnamefont
  {Essig}},\ }\href {\doibase 10.1103/PhysRevD.78.015004} {\bibfield  {journal}
  {\bibinfo  {journal} {Phys. Rev.}\ }\textbf {\bibinfo {volume} {D78}},\
  \bibinfo {pages} {015004} (\bibinfo {year} {2008})},\ \Eprint
  {http://arxiv.org/abs/0710.1668} {arXiv:0710.1668 [hep-ph]} \BibitemShut
  {NoStop}%
\bibitem [{\citenamefont {Elor}\ \emph {et~al.}(2016)\citenamefont {Elor},
  \citenamefont {Rodd}, \citenamefont {Slatyer},\ and\ \citenamefont
  {Xue}}]{Elor:2015bho}%
  \BibitemOpen
  \bibfield  {author} {\bibinfo {author} {\bibfnamefont {G.}~\bibnamefont
  {Elor}}, \bibinfo {author} {\bibfnamefont {N.~L.}\ \bibnamefont {Rodd}},
  \bibinfo {author} {\bibfnamefont {T.~R.}\ \bibnamefont {Slatyer}}, \ and\
  \bibinfo {author} {\bibfnamefont {W.}~\bibnamefont {Xue}},\ }\href {\doibase
  10.1088/1475-7516/2016/06/024} {\bibfield  {journal} {\bibinfo  {journal}
  {JCAP}\ }\textbf {\bibinfo {volume} {1606}},\ \bibinfo {pages} {024}
  (\bibinfo {year} {2016})},\ \Eprint {http://arxiv.org/abs/1511.08787}
  {arXiv:1511.08787 [hep-ph]} \BibitemShut {NoStop}%
\bibitem [{\citenamefont {Profumo}\ \emph {et~al.}(2018)\citenamefont
  {Profumo}, \citenamefont {Queiroz}, \citenamefont {Silk},\ and\ \citenamefont
  {Siqueira}}]{Profumo:2017obk}%
  \BibitemOpen
  \bibfield  {author} {\bibinfo {author} {\bibfnamefont {S.}~\bibnamefont
  {Profumo}}, \bibinfo {author} {\bibfnamefont {F.~S.}\ \bibnamefont
  {Queiroz}}, \bibinfo {author} {\bibfnamefont {J.}~\bibnamefont {Silk}}, \
  and\ \bibinfo {author} {\bibfnamefont {C.}~\bibnamefont {Siqueira}},\ }\href
  {\doibase 10.1088/1475-7516/2018/03/010} {\bibfield  {journal} {\bibinfo
  {journal} {JCAP}\ }\textbf {\bibinfo {volume} {1803}},\ \bibinfo {pages}
  {010} (\bibinfo {year} {2018})},\ \Eprint {http://arxiv.org/abs/1711.03133}
  {arXiv:1711.03133 [hep-ph]} \BibitemShut {NoStop}%
\bibitem [{\citenamefont {Leane}\ \emph {et~al.}(2018)\citenamefont {Leane},
  \citenamefont {Slatyer}, \citenamefont {Beacom},\ and\ \citenamefont
  {Ng}}]{Leane:2018kjk}%
  \BibitemOpen
  \bibfield  {author} {\bibinfo {author} {\bibfnamefont {R.~K.}\ \bibnamefont
  {Leane}}, \bibinfo {author} {\bibfnamefont {T.~R.}\ \bibnamefont {Slatyer}},
  \bibinfo {author} {\bibfnamefont {J.~F.}\ \bibnamefont {Beacom}}, \ and\
  \bibinfo {author} {\bibfnamefont {K.~C.~Y.}\ \bibnamefont {Ng}},\ }\href
  {\doibase 10.1103/PhysRevD.98.023016} {\bibfield  {journal} {\bibinfo
  {journal} {Phys. Rev.}\ }\textbf {\bibinfo {volume} {D98}},\ \bibinfo {pages}
  {023016} (\bibinfo {year} {2018})},\ \Eprint
  {http://arxiv.org/abs/1805.10305} {arXiv:1805.10305 [hep-ph]} \BibitemShut
  {NoStop}%
\bibitem [{\citenamefont {Cline}\ \emph {et~al.}(2014)\citenamefont {Cline},
  \citenamefont {Dupuis}, \citenamefont {Liu},\ and\ \citenamefont
  {Xue}}]{Cline:2014dwa}%
  \BibitemOpen
  \bibfield  {author} {\bibinfo {author} {\bibfnamefont {J.~M.}\ \bibnamefont
  {Cline}}, \bibinfo {author} {\bibfnamefont {G.}~\bibnamefont {Dupuis}},
  \bibinfo {author} {\bibfnamefont {Z.}~\bibnamefont {Liu}}, \ and\ \bibinfo
  {author} {\bibfnamefont {W.}~\bibnamefont {Xue}},\ }\href {\doibase
  10.1007/JHEP08(2014)131} {\bibfield  {journal} {\bibinfo  {journal} {JHEP}\
  }\textbf {\bibinfo {volume} {08}},\ \bibinfo {pages} {131} (\bibinfo {year}
  {2014})},\ \Eprint {http://arxiv.org/abs/1405.7691} {arXiv:1405.7691
  [hep-ph]} \BibitemShut {NoStop}%
\end{thebibliography}%

\end{document}